\documentclass[journal,draftclsnofoot,onecolumn,12pt,twoside]{IEEEtran}
\usepackage{soul}
\usepackage{graphicx}
\usepackage{amsmath}
\usepackage{amsthm}
\usepackage{amsfonts}
\usepackage[justification=centering]{caption}
\usepackage{color, xcolor}
\usepackage{diagbox}
\usepackage{booktabs}
\usepackage{multirow}
\usepackage{algorithm}
\usepackage{algpseudocode}
\usepackage{amssymb}
\usepackage{cite}
\usepackage{hyperref} 
\usepackage{blkarray}
\usepackage{caption}
\usepackage{subcaption}
\usepackage{pstricks}
\usepackage{rotating}
\usepackage{booktabs} 
\usepackage{ulem}
\usepackage{blkarray} 
\allowdisplaybreaks

\theoremstyle{}
\newtheorem{theorem}{Theorem}
\newtheorem{lemma}{Lemma}

\newtheorem{corollary}{Corollary}

\newtheorem{example}{Example}
\newtheorem{remark}{Remark}

\newcommand{\tabcaption}{\def\@captype{table}\caption}
\newcommand{\tabincell}[2]{\begin{tabular}{@{}#1@{}}#2\end{tabular}}

\makeatletter
\newcommand*\bigcdot{\mathpalette\bigcdot@{.5}}
\newcommand*\bigcdot@[2]{\mathbin{\vcenter{\hbox{\scalebox{#2}{$\m@th#1{\bullet}$}}}}}

\makeatother

\normalsize
\ifCLASSINFOpdf

\else

\fi

\begin{document}
\title{Order Optimal Cascaded Coded Distributed Computing With Low Complexity and Improved Flexibility
\author{\IEEEauthorblockN{Mingming Zhang, Youlong Wu, Minquan Cheng, and Dianhua Wu}
	\thanks{M. Zhang, M. Cheng and D. Wu are with Guangxi Key Lab of Multi-source Information Mining $\&$ Security, Guangxi Normal University,
		Guilin 541004, China  (e-mail: ztw\_07@foxmail.com, chengqinshi@hotmail.com), dhwu@gxnu.edu.cn.}
	\thanks{Y. Wu is with the School of Information Science and Technology, ShanghaiTech University, Shanghai 201210, China  (e-mail:  wuyl1@shanghaitech.edu.cn).}
}
}
\maketitle

\begin{abstract}
Coded distributed computing (CDC), proposed by Li \emph{et al.}, offers significant potential for reducing the communication load in MapReduce computing systems.
{ In CDC with $K$ nodes, $N$ input files, and $Q$ output functions, each input file will be mapped by $r\geq 1$ nodes and each output function will be computed by $s\geq 1$ nodes such that coding techniques can be applied to generate multicast opportunities.} 
{ However, a significant limitation of most existing coded distributed computing schemes is their requirement to split the original data into a large number of input files (or output functions) that grows exponentially with $K$, which significantly increases the coding complexity and degrades the system performance.} 
In this paper, we focus on the case of $K/s\in\mathbb{N}$, deliberately designing the strategy of data placement and output functions assignment, such that a low-complexity CDC scheme is achievable.
The main advantages of the proposed scheme include: 1) the multicast gains equal to $(r+s-1)(1-1/s)$ and $r+s-1$ { which is approximately} $r+s-1$ when $s$ is relatively large, { and the communication load potentially better than the well-known scheme proposed by Li \emph{et al.};} 2) the proposed scheme requires significantly { less input files} and output functions; 3) all the operations are implemented over the binary field $\mathbb{F}_2$ with the one-shot fashion { (i.e., each node can decode its requested content immediately upon receiving the multicast message during the current time slot)}.
Finally, we derive a new information-theoretic converse bound for the cascaded CDC framework under the proposed strategies of data placement and output functions assignment.
We demonstrate that  the communication load of the proposed scheme is order optimal within a factor of $2$; and is also approximately optimal when $K$ is sufficiently large for a given $r$.
\end{abstract}


\section{Introduction}
Distributed computing systems have been extensively employed for the execution of large-scale computing tasks due to their ability to significantly enhance execution speed.
This is achieved by allowing computation jobs to run in parallel while leveraging distributed computing and storage resources. However, in the context of massive data exchanged among computing nodes, distributed computing systems face a severe communication bottleneck arising from limited resources and high traffic load. For instance, in applications such as TeraSort \cite{TeraSort} and SelfJoin \cite{SelfJoin} running on an Amazon EC2 cluster, the data exchange process costs $65\%\sim 70\%$ of the overall job execution time \cite{CZMJS}.

In the pioneering paper \cite{LMYA}, Li \emph{et al.} extended the coding technique to distributed computing and proposed the coded distributed computing (CDC) scheme, in order to reduce the communication load by increasing the computations of input data. 
{ In the considered MapReduce setting  with $K$ distributed computing nodes, $N$ input files, and $Q$ output functions, each input file will be mapped by $r$ nodes and each output function will be computed by $s$ nodes. Here $r$ represents the computation load\footnote{In \cite{LMYA}, the computational load is defined as the total number of computed Map functions at the nodes, normalized by $N$. This paper focuses on the symmetric scenario where each file is mapped by exactly $r$ workers.}.}When { $s>1$}, the setting is called  \textit{cascaded coded distributed computing (cascaded CDC)}, which matches the multiple rounds distributed computing jobs where the output results of the previous round serve as the input of the next round. 
{ The cascaded CDC contains “Map”, “Shuffle” and “Reduce” phases. In the Map phase, each node stores and maps the local input files, generating $Q$ intermediate values (IVs) of equal size $T$-bit for each input file. In the Shuffle phase, the nodes generate coded messages from the local IVs, and multicast them to other nodes, ensuring that all necessary IVs are recoverable by the desired nodes. In the Reduce phase, each output function will be reduced by the assigned $s$ nodes based on the locally computed and recovered IVs. The end-goal of the computation is to compute the output functions in a distributed manner. Denote the communication load to be the total communication bits normalized by $NQT$. }
In \cite{LMYA}, the authors demonstrated that their cascaded scheme achieves optimal communication load when  $N/\binom{K}{r} \in\mathbb{N}$\footnote{The condition $N/\binom{K}{r} \in\mathbb{N}$  can be relaxed if $N$ is sufficiently large.} and  $Q/\binom{K}{s} \in\mathbb{N}$,  
and the multicast gains { (the average number of nodes served by one multicast message during the Shuffle phase)} in multiple rounds transmission equal to $r$, $r+1$, $\ldots$, $\min\{r+s-1, K-1\}$.
Following the aforementioned CDC system, various problems have been studied in \cite{CDC1,YWYT,JC,horii2020improved,KR,compresscoded,WJC,chen2020coded,woolsey2021new,xu2021new,wang2021batch,reisizadeh2019coded,kim2019optimal,WWCJ,ELMA,YYW,Wang22,CDCsurvey,BE,BWCE,CLMTW}. 

\subsection{Research Motivation}
\label{subsec-research motivation}
{ The above-mentioned cascaded CDC scheme needs exponentially large quantities of both input files and output functions concerning $K$, which are increased with  $\binom{K}{r}$ and $\binom{K}{s}$, respectively.
	As a consequence, practical implementations encounter unexpected performance degradation, particularly when $K$ assumes relatively large values.  
	To reduce the required numbers of both input files and output functions, several schemes with the lower-complexity are proposed in \cite{JQ,WCJ,JWZ,CWL,CLCEL,CLCEL2}, which are detailed in Section~\ref{subsec-performance analysis}.}
However, these schemes impose stringent requirements on system parameters $(r,s, K,N,Q)$, which poses obstacles to the implementation of existing schemes in practical scenarios. More specifically,
\begin{itemize}
	\item In \cite{WCJ,JWZ,CWL,CLCEL,CLCEL2}, the computation load $r$ and the number of computations per output function $s$ exhibit a strong correlation, i.e., $r=s$ or $r+s=K$. 
	However, in practice, $r$ is the computation load assigned to the system that should depend on the hardwares or the trade-off level of communication-computation, while   $s$ is only associated with the computation tasks and    is  often independent of $r$.
	
	\item In \cite{WCJ,JWZ,CWL,CLCEL,CLCEL2}, the numbers of input files $N$ and output functions $Q$ should be equal. This means that when implementing the existing schemes, the system should split the data into $N$ files according to the number of output functions $Q$, i.e., $N=Q$. If $Q$ is relatively small compared to $N$, the previous schemes will no longer work as the number of files are not enough to support the repetitive data placement in their schemes.
	
	
	\item In \cite{JWZ,CWL,CLCEL,CLCEL2}, the feasible number of computing nodes $K$ is specific and takes the form of combinations, or as a polynomial of primes. This indicates that the number of computation nodes in their schemes  should be sufficiently large, and is not feasible for small-scale distributed  computing schemes. 
\end{itemize}

Furthermore, only the schemes in \cite{JQ} and \cite{WCJ} can support for flexible $K$ among all existing schemes reducing coding complexity in terms of  $(N,Q)$.
It is interesting that  these two schemes both are designed based on the grouping approach from the view point of combinatorial design. More precisely, both \cite{JQ} and \cite{WCJ}   adopt the same  data placement strategy based on the combinatorial structure named Partition or hypercube. The difference between \cite{JQ} and \cite{WCJ}  lies on the  output functions assignment  and Shuffle phase delivery strategies,  inducing their individual strengths and weaknesses:

\begin{itemize}
	\item In \cite{JQ}, the output function assignment of each group is the same. 	
	The number of required output functions $Q$ is linearly increasing with $K$, and the scheme operates in a binary field, $\mathbb{F}_2$.
	However, it fails to effectively utilize common IVs among multiple nodes and leads to redundant communication load, where the multicast gain is only $(r-1)/s$.
	
	\item In \cite{WCJ}, the output function assignment is the same as its file storage. During the Shuffle phase, each node multicasts linear combinations of IVs segments, and the resulting scheme with low communication load is proved to be order optimal under the specific output function assignment.  \textit{Particularly, when $r=s=2$, the communication load is strictly less than that of \cite{LMYA}}, where the multicast gain is $\frac{3K}{2(K+1)}$. However, the value of $Q$ still grows exponentially with $K$, and the scheme is not feasible in a binary field.
\end{itemize}

In short, the   scheme in \cite{JQ} demands a small number of output functions $Q$ and operates in a binary field, but it suffers from poor communication load performance, while  the scheme in \cite{WCJ}  exhibits low communication load, but   still requiring $Q$  to  grow exponentially with $K$ and works in large finite fields. 
\emph{For the CDC scheme flexible in $K$, an interesting question is whether one can achieve low communication load approaching to or even better than \cite{WCJ}, while simultaneously keeping small $N,Q$ and working in binary field with low complexity. }

In this paper, we give an affirmative answer by proposing   communication-efficient, flexible, and low-complexity cascaded CDC scheme that reduce both the communication load and required values of  $(N,Q)$, while breaking the stringent requirements in existing works, including 1) $r=s$ or $r+s=K$; 2) $N=Q$; 3) $K$   takes the form of combinations or as a polynomial of primes.  
{ In Table~\ref{tab-limit}, we summarize the limitations of lower-complexity schemes in \cite{JQ,WCJ,JWZ,CWL,CLCEL,CLCEL2}, and compare them with the proposed scheme.}
\begin{table}
	\centering
	\caption{Summary of limitations in existing cascaded CDC schemes compared to the proposed scheme}\label{tab-limit}
	\resizebox{1.0\linewidth}{!}{
	\begin{tabular}{|c|c|c|c|c|c|}
		\hline
		Schemes & Flexibility in $K$ & Flexibility in $r, s$ & Flexibility in $N, Q$ & Communication-efficient & Operation  field $\mathbb{F}_2$ \\ \hline
		\cite{LMYA} & $\checkmark$ & $\checkmark$ & $\checkmark$ & $\checkmark$ & $\times$ \\ \hline
		\cite{JQ} & $\checkmark$ & $\checkmark$ & $\checkmark$ & $\times$ & $\checkmark$ \\ \hline
		\cite{WCJ} & $\checkmark$ & $\times$ & $\times$ & $\checkmark$ & $\times$ \\ \hline
		\cite{JWZ} & $\times$ & $\times$ & $\times$ & $\checkmark$ & $\times$ \\ \hline
		\cite{CWL} & $\times$ & $\times$ & $\times$ & $\checkmark$ & $\checkmark$ \\ \hline
		\cite{CLCEL} & $\times$ & $\times$ & $\times$ & $\checkmark$ & $\times$ \\ \hline
		\cite{CLCEL2} & $\times$ & $\times$ & $\times$ & $\checkmark$ & $\times$ \\ \hline
		\hline
		Proposed scheme & $\checkmark$ & $\checkmark$ & $\checkmark$ & $\checkmark$ & $\checkmark$ \\ \hline 
	\end{tabular}}
\end{table}

\subsection{Contributions and Paper Organization}
\label{subsec-contribution}
In this paper, we propose a novel cascaded CDC scheme that reduces the communication load for the MapReduce system with low computational complexity and more flexible system parameters  $(r,s, K,N,Q)$. Our main contributions are summarized as follows.

\begin{itemize}


\item Unlike the existing lower-complexity schemes in \cite{WCJ,JWZ,CWL,CLCEL,CLCEL2} requiring   $r\in\{s,K-s\}$ and $N=Q$,  we propose a novel CDC scheme that not only supports   flexible setting   with  $K/s\in\mathbb{N}$ and $N$ being independent of $Q$, but also greatly reduces the communication load. 
{ We deliberately design the strategies of data placement, output functions assignment, and two-rounds Shuffle delivery, the achieved multicast gains are $(r+s-1)(1-1/s)$ and $r+s-1$, respectively.} 
When $s$ is relatively large, the multicast gain is approximate to $r+s-1$, which is  the maximum under one-shot linear delivery  \cite{CWL}.

\item  { The proposed scheme shows that using new  strategies of data placement and output functions assignment can potentially break the fundamental limits (under a given assignment of output functions) presented in \cite{LMYA}. 
	In prior works with flexible in $K$, only \cite{WCJ} broke the limit under the specific condition $r=s=2$. 
	Our work reveals that even for $r\neq s$,  there exist numerous cases where the limits in \cite{LMYA} can still be broken.}

%


\item The proposed scheme has low complexity in terms of small $(N,Q)$ \footnote{In CDC schemes, large $N$ and $Q$ will incur large computational complexity \cite{LMYA}.} and one-shot binary delivery. { Specifically,   in our scheme  the  number of input files $N$  and output functions $Q$ are increased with ${K/s\choose (r-1)/s+1}(r+s-1)$ and $(K/s)$, instead of ${K\choose r}$ and ${K\choose s}$ in \cite{LMYA},} which is significantly relaxing the requirements for exponentially large numbers of input files and output functions. 
{ Compared to the lower-complexity schemes with flexible $K$ \cite{JQ,WCJ}, our scheme also achieves an exponential reduction in $(N,Q)$.} 
Moreover, the proposed one-shot transmission strategy during the Shuffle phase operates on the binary field $\mathbb{F}_2$, significantly reducing the computational complexity.

\item We derive a new information-theoretic converse bound for the cascaded CDC framework, under the given strategies of data placement and output functions assignment.
Furthermore, we prove that the proposed scheme achieves the \textit{order optimal} communication load that is within a $2$ factor of the information-theoretic optimum; and for any given $r$, when $K\to\infty$, it is \textit{asymptotically optimal}.
We also have the optimality under the constraint of $s=1$, consistent with the result in \cite{LMYA}.

\end{itemize}

This paper is organized as follows. In Section~\ref{sec-system model}, the system model and problem formulation are introduced. In Section~\ref{sec-main result},  main results and  performance analysis are proposed. An { illustrative example} of the proposed achievable scheme is shown in Section~\ref{sec-example}. The detailed constructions of the proposed scheme are stated in Section~\ref{sec-proof}. Finally, Section~\ref{sec-conclusion} concludes this paper, and Appendices provide some proofs.

\subsection*{Notations}
In this paper,   the following notations are used  unless otherwise stated.
\begin{itemize}
\item Bold capital letter, bold lower case letter and curlicue font are used to denote array { (refers to a multidimensional structure)}, vector { (refers to a one-dimensional structure)} and set respectively.
$|\cdot|$ is the length of a vector  or the cardinality of a set;
	
\item For any positive integers $a$, $b$, $t$ with $a<b$ and $t\leq b $,   non-negative set $\mathcal{V},\mathcal{J}$,
	\begin{itemize}
		\item   $[a]:=\{1,2,\ldots,a\}$, $[a:b] :=\{a,a+1,\ldots,b\}$, $[a:b):=\{a,a+1,\ldots,b-1\}$ and ${[b]\choose t}:=\{\mathcal{V}\ :\   \mathcal{V}\subseteq [b], |\mathcal{V}|=t\}$, i.e., ${[b]\choose t}$ is the collection of all subsets of $[b]$ with cardinality $t$. Let $a|b$ denote  that $b/a\in\mathbb{N}$.

		\item { The elements of each set are listed in increasing order unless otherwise specified.}
		$\mathcal{V}[j]$ represents the $j^{\text{th}}$ smallest element of $\mathcal{V}$, where $j\in[|\mathcal{V}|]$, and $\mathcal{V}[\mathcal{J}]=\{\mathcal{V}_j:j\in\mathcal{J}\}$.
	\end{itemize}

\end{itemize}

\section{System Model}
\label{sec-system model}
{ As formulated in \cite{LMYA}, the end goal of the CDC model is to compute  $Q$ output functions from $N$ input files using $K$ distributed computing nodes.}
The $N$ input files are denoted by $w_1,\ldots,w_N$;
the $Q$ output functions are denoted by $\phi_1,\ldots,\phi_Q$.
For each $q\in[Q]$,  define $\mathcal{A}_q\subseteq[K]$ to be the set of nodes that are assigned to compute the output function $\phi_q$\footnote{{ For a CDC scheme, $\mathcal{A}_q$ indicates the strategy of \textit{output function assignment}. 
In other words, each output function $\phi_q$, where $q\in[Q]$, is assigned to the subset of nodes $\mathcal{A}_q\subseteq[K]$.}}, which maps all the $N$ input files into a stream
\begin{align}
	u_q&\triangleq\phi_q(w_{1},w_{2},\ldots,w_{N})\\
	&\triangleq h_q(g_{q,1}(w_{1}),g_{q,2}(w_{2}),\ldots,g_{q,N}(w_{N})) \in \mathbb{F}_{2^E},
\end{align}
for some integer $E$. For any $q\in [Q]$ and $n\in [N]$,  $g_{q,n}(\cdot)$ is  called Map function and $h_q(\cdot)$ is called Reduce function. The parameter $v_{q,n}\triangleq g_{q,n}(w_{n})$ where $q\in [Q]$ and $n\in [N]$ is called intermediate value (IV). Each IV has $T$ bits for some positive $T$. 
{ We assume that each output function is assumed to be computed by $s\in[K]$ nodes and each input file will be mapped by $r$ nodes. In cascaded computing, the output values will be taken as the input of the next round. Therefore, by letting each output function be computed by $s$ nodes, it will help reduce the communication overhead.}
The  CDC scheme consists of the following three phases.

\textit{Map Phase}:
For each $n\in[N]$,  define $\mathcal{D}_n\subseteq[K]$ as the set of nodes that  placed  the input file $w_n$\footnote{{ For a CDC scheme, $\mathcal{D}_n$ indicates the strategy of \textit{data placement}. 
In other words, each input file $w_n$ where $n\in[N]$, is placed at the subset of nodes $\mathcal{D}_n\subseteq[K]$. }}
Thus for each node  $k\in[K]$, the set of stored input files is denoted by
\begin{align}
\mathcal{M}_k=\{w_n:n\in[N],k\in\mathcal{D}_n\}. \label{eq-model Mk}
\end{align}
{ Similar to the existing works \cite{LMYA,JQ,WCJ,JWZ,CWL,CLCEL,CLCEL2}, we also assume that each node maps the same number of files, say $M$ files, i.e., $|\mathcal{M}_k|=M$ for each $k\in[K]$. 
Since each file is mapped by $r\in[K]$ nodes, we have $Nr=KM$.}
By using the stored files in \eqref{eq-model Mk} and Map functions $\{g_{q,n}(\cdot)\}$ where $q\in[Q]$ and $n\in[N]$, each node $k$ computes the IVs $\mathcal{I}_k=\{v_{q,n}=g_{q,n}(w_{n})\ |\ q\in [Q],w_n\in \mathcal{M}_k\}$ locally.
{ By the end of Map phase,  the IVs for all $Q$ output functions are generated across the $K$ nodes. Thus there are $N\times Q$ different IVs overall.}

\textit{Shuffle Phase}:
According to the output functions assignment, for each node $k\in[K]$, the set of assigned output functions is denoted by
\begin{align}
\mathcal{W}_k=\{\phi_q:q\in[Q],k\in\mathcal{A}_q\}. \label{eq-model Wk}
\end{align}
Each node transmits multicast message(s) derived from locally computed IVs to the other nodes, such that after the Shuffle phase, each node can recover all necessary IVs of the assigned output functions.
We denote the multicast message(s) transmitted by node $k\in[K]$ as $X_k\in\mathbb{F}_{2^{l_k}}$.
{ Having generated the message $X_k$, node $k$ multicasts it to all other nodes. By the end of Shuffle phase, each of the $K$ nodes receives $X_1,\ldots,X_K$ free of error.}

\textit{Reduce Phase}:
During the Reduce phase, each node $k\in[K]$ reconstructs all the needed IVs for each $q\in\mathcal{W}_k$ by using the messages transmitted in the Shuffle phase and the IVs computed locally in $\mathcal{I}_k$.

The \textit{computation load} is defined as the normalized total number of files stored by all the $K$ nodes, which is formulated by
\begin{align}
	r:=\frac{\sum_{k\in[K]}|\mathcal{M}_k|}{N}\label{eq-def of r}.
\end{align}

The \textit{communication load} is defined as the normalized total number of bits transmitted by all the $K$ nodes during the Shuffle phase, which is formulated by
\begin{align}
L:=\frac{\sum_{k\in[K]}|X_k|}{QNT}=\frac{\sum_{k\in[K]}l_k}{QNT}\label{eq-def of L}.
\end{align}

When $N/{K\choose r}\in\mathbb{N}$ and $Q/{K\choose s}\in\mathbb{N}$  (or $N\rightarrow\infty,Q\rightarrow\infty$), in \cite{LMYA} Li \textit{et al.} proposed the first well-known CDC scheme achieving the minimum communication load under the specific strategies of data placement and output function assignment, which is illustrated as follows.
\begin{lemma}[\cite{LMYA}]
For any positive integers $K$, $r$ and $s$, there exists a cascaded CDC scheme achieving the minimum communication load $L_{\text{Li \textit{et al.}}}(r,s)$ under its given strategies of data placement and output function assignment,
\begin{align}\label{eq-converse}
	L_{\text{Li \textit{et al.}}}(r,s)=\sum\limits_{l=\max\{r+1,s\}}^{\min\{r+s,K\}}\left(\frac{{K-r\choose K-l}{r\choose l-s}}{{K\choose s}}\cdot \frac{l-r}{l-1}\right),
\end{align}
where $r$ is the computation load,  $s$ is the number of nodes that compute each reduce function.
\end{lemma}

In \cite[Lemma 7]{CWL}, the authors provided the maximum multicast gain under the one-shot linear delivery.
\begin{lemma}[\cite{CWL}]
\label{lem-multicasting gain}
For the $(K,r,s,N,Q)$ cascaded CDC scheme under one-shot linear delivery, the maximum multicast gain $g^*_{\text{one-shot}}\leq\min\{r+s-1,K-1\}$.
\end{lemma}

We aim to design a low-complexity and communication-efficient scheme that reduces the communication load $L$ for given $r$ and $s$, with small quantities of $N$ and $Q$.

\section{Main results}
\label{sec-main result}

By deliberately designing the input file and output function assignments, we propose the cascaded CDC scheme as follows. The proof could be found in Section~\ref{sec-proof}.
\begin{theorem}[Achievable Scheme]
\label{th-proposed scheme}
In the cascaded CDC framework with $s>1$ and $K/s\in\mathbb{N}$, the following  communication load is achievable,
\begin{align}
L(r,s)=\frac{s(K-r+1)}{K(r+s-1)}, \ \ \ \forall r\in\{1,s+1,2s+1,\ldots,(K/s-1)s+1\},
\label{eq-Load}
\end{align}
and $L(K,s)=0$.

{ The proposed scheme requires that the number of input files is a multiple of $\left({K/s\choose (r-1)/s+1}(r+s-1)\right)$, and the number of output functions is a multiple of $(K/s)$.}

\end{theorem}

\begin{corollary}
\label{corol-multicasting gain}
For the scheme in Theorem~\ref{th-proposed scheme}, the { multicast gain} is consisted of $(r+s-1)(1-1/s)$ and $r+s-1$ (detailed by \eqref{eq-g1} and \eqref{eq-g2} in Section~\ref{sec-proof}).
By Lemma~\ref{lem-multicasting gain}, the scheme achieves one-shot asymptotically optimal communication load if $s\to\infty$.
\end{corollary}

By exploiting the method in \cite{KNL}, we derive a new information theoretic converse bound for the cascaded CDC framework with $K/s\in\mathbb{N}$, under the given strategies of data placement and output functions assignment.
The proposed scheme is proved to be order optimal within a factor of 2; and for a given $r$, when $K\to\infty$, it is proved to be asymptotically optimal.
The proof of Theorem~\ref{th-converse} could be found in Appendix~\ref{app-converse}.

\begin{theorem}[Converse Bound]
\label{th-converse}
In the cascaded CDC framework with $s>1$ and $K/s\in\mathbb{N}$,  { under the data placement and output function assignment strategies of the proposed CDC scheme}, the converse bound satisfies that
\begin{align}
L^{\star}(r,s)\geq\frac{s(s-1)}{K(r+s-2)}+\frac{s(K-(r+s-1))}{K(r+s-1)}, \ \ \ \forall r\in\{1,s+1,2s+1,\ldots,(K/s-1)s+1\},
\label{eq-load star}
\end{align}
and $L^{\star}(K,s)=0$.
The proposed scheme is order optimal within a factor of 2.
Furthermore, for a given $r$, when $K\to\infty$, the proposed scheme is asymptotically optimal.
\end{theorem}

{
\subsection{Performance Analysis}
\label{subsec-performance analysis}
Next, we will show the advantages of the proposed scheme compared to the lower-complexity schemes in \cite{JQ,WCJ,JWZ,CWL,CLCEL,CLCEL2}. Prior to the comparisons, we first introduce these schemes.
\begin{itemize}
	\item In \cite{JQ}, the authors generated cascaded CDC schemes by using the Partition placement delivery array (which is originally proposed in \cite{YCTC}) under the binary field $\mathbb{F}_2$.
	However,  these schemes fail to effectively utilize the common IVs required by multiple nodes, resulting in redundant communication load.
	\item In \cite{WCJ}, the authors proposed a scheme based on hypercube structure under the constraint of $r=s$ with { fewer} input files and output functions, and a high-performance communication load that is asymptotically optimal.
	The authors also showed that when $r = s = 2$, the scheme can achieve a strictly smaller communication load than that the scheme in \cite{LMYA}. 
	However, both the number of input files and output functions are still exponentially in the form of $(\frac{K}{r})^{r}$.
	\item In \cite{JWZ}, the authors employed a symmetric balanced incomplete block design (SBIBD) to present a strategy of  data placement and output functions assignment,  ultimately generating an asymptotically cascaded scheme with $N=Q=K$, while under the constraint of $r=s$ or $r+s=K$.
	\item In \cite{CWL}, the authors proved that the maximum multicast gain under the one-shot linear delivery{ \footnote{ { Focusing on Shuffle phase, we refer to a delivery scheme as one-shot linear delivery if each multicast message is a linear combination of IVs and can be directly decoded by the desired nodes in one-round communication with a linear operation.}}} is $r+s-1$.
	When $r=s$, the low-complexity and asymptotically optimal cascaded schemes are proposed  under the minimum binary field $\mathbb{F}_2$, by using combinatorial structures $t$-design ($t\geq2$) and $t$-group divisible design ($t$-GDD), respectively. 
	Based on \cite{CWL}, the schemes in \cite{CLCEL} and \cite{CLCEL2} are proposed by using some other combinatorial design structure.
	\item In \cite{CLCEL}, the authors proposed two asymptotically optimal cascaded schemes with $N=Q=K$. One of them used symmetric design (SD) for the case of $r+s=K$, while achieving a lower communication load than that of \cite{JWZ}; the other one uses almost difference (AD) sets.
	\item In \cite{CLCEL2}, focusing on the case of $r=s$, the authors proposed an improved asymptotically optimal cascaded scheme relying on another special $t$-design.
	However, from \cite{CD}, the AD sets and the specific $t$-design in \cite{CLCEL} and \cite{CLCEL2} are rare at present.
\end{itemize}
These schemes are summarized in Table~\ref{tab-comp}\footnote{{ Table~\ref{tab-comp} gives the smallest number of $N$ (the number of input files) and $Q$ (the number of output functions). For each scheme, the values of $N$ and $Q$ can be chosen as integer multiples of the corresponding entries in the table. For instance, in \cite{LMYA} the required $N$ and $Q$ satisfy $N/{K\choose r}\in\mathbb{N}$ and $Q/{K\choose s}\in\mathbb{N}$, respectively.}}. 
Compared with the existing cascaded CDC schemes, the advantages of the proposed scheme include: 1) communication-efficient; 2) keeping a small number of $N, Q$; 3) breaking the stringent parameter requirements in existing works\footnote{{ The stringent parameter requirements include 1) $r=s$ or $r+s=K$; 2) $N=Q$; 3) $K$ takes the form of combinations or as a polynomial of primes.}}; and 4) all the operations are implemented over the binary field $\mathbb{F}_2$ with the one-shot fashion.

\begin{table}
	\centering
	\vspace{-0.8cm}
	\caption{Comparisons of existing cascaded CDC schemes and the proposed scheme }\label{tab-comp}
	\begin{tabular}{|c|c|c|c|c|c|c|c|}
		\hline
		Parameters & \tabincell{c}{Number of \\ nodes $K$} & \tabincell{c}{Computation \\ load $r$} & \tabincell{c}{Replication \\ factor $s$} & \tabincell{c}{Number \\ of input \\ files $N$} & \tabincell{c}{Number\\ of  output \\ functions $Q$} & \tabincell{c}{Communication \\ load $L$} & \tabincell{c}{Operation \\ field $\mathbb{F}_2$} \\ \hline
		\tabincell{c}{\cite{LMYA}: \\ $K,r,s\in\mathbb{N}$,\\ $1\leq r$,\\ $1\leq s\leq K$} & $K$ & $r$ & $s$ & ${K\choose r}$ & ${K\choose s}$ & \tabincell{c}{$\sum\limits_{l=\max\{r+1,s\}}^{\min\{r+s,K\}}$\\ $\left(\frac{{K-r\choose K-l}{r\choose l-s}}{{K\choose s}}\cdot\frac{l-r}{l-1}\right)$} & No \\ \hline
		\tabincell{c}{\cite{JQ}:\\ $K,r,s\in\mathbb{N}$,\\ $K/r\in\mathbb{N}$,\\ $K/\text{gcd}(K,s)$\\ $\in\mathbb{N}$} & $K$ & $r$ & $s$ & $(\frac{K}{r})^{r-1}$ & $\frac{K}{\text{gcd}(K,s)}$ & $\min\{\frac{s(1- {r}/{K})}{r-1},1\}$ &  Yes \\ \hline
		\tabincell{c}{\cite{WCJ}:\\ $K,r\in\mathbb{N}$,\\ $K/r\in\mathbb{N}$} & $K$ & $r$ & $r$ & $(\frac{K}{r})^{r}$ & $(\frac{K}{r})^{r}$ & \tabincell{c}{$\frac{1}{2}-\frac{1}{2}\cdot(\frac{r}{K})^r+$\\ $\frac{1}{4r-2}\cdot(1-\frac{r}{K})^r$}  & No \\ \hline
		\multirow{4}{*}{\shortstack{\tabincell{c}{\cite{JWZ}: $b$ is a\\prime power}} } & $b^2+b+1$ & $b^2$ & $b^2$ & $K$ & $K$ & $\frac{r}{r-1}(1-\frac{r}{K})$ & \multirow{4}{*}{No}\\ \cline{2-7}
		& $b^2+b+1$ & $b+1$ & $b^2$ & $K$ & $K$  & \multirow{3}{*}{$\frac{K}{K-1}(1-\frac{r}{K})$} & \\ \cline{2-6}
		& $b^3+b^2+b+1$ & $b^2+b+1$ & $b^3$ & $K$ & $K$  &  &  \\ \cline{2-6}
		& $b^3+2b^2$ & $b^2+b$ & $b^3+b^2-b$ & $K$ & $K$  &  &  \\ \hline
		\tabincell{c}{\cite{CWL}:\\ $t$-$(N,M,\lambda)$\\ design} & $\frac{\lambda{N\choose t}}{{M\choose t}}$  & $\frac{KM}{N}$ & $\frac{KM}{N}$ & $N$ & $N$ & $\frac{N-1}{2N}<\frac{1}{2}$ & \multirow{2}{*}{Yes}  \\ \cline{1-7}
		\tabincell{c}{\cite{CWL}:\\ $t$-\\$(m,q,M,\lambda)$\\ GDD} & $\frac{\lambda{m\choose t}q^t}{{M\choose t}}$  & $\frac{KM}{mq}$ & $\frac{KM}{mq}$ & $mq$ & $mq$ & $\frac{1}{2}+\frac{q-2}{2mq}$ &   \\ \hline
		\multirow{3}{*}{\shortstack{\cite{CLCEL}:  $b$ is a\\prime power} } & $b^2+b+1$ & $b+1$ & $b^2$ & $K$ & $K$  & \multirow{3}{*}{$\frac{(K-1)^2-rK+K}{K(K-1)}$} & \multirow{5}{*}{No} \\ \cline{2-6}
		& $b^3+b^2+b+1$ & $b^2+b+1$ & $b^3$ & $K$ & $K$  &  &  \\ \cline{2-6}
		& $b^3+2b^2$ & $b^2+b$ & $b^3+b^2-b$ & $K$ & $K$  &  &  \\ \cline{1-7}
		\tabincell{c}{\cite{CLCEL}:\\ $(n,k,\lambda,\mu)$\\ AD set with\\ $ \lambda<k-1$} & $n$ & $k$ & $k$ & $K$ & $K$ & $\frac{n-1}{2n}$ & \\ \cline{1-7}
		\tabincell{c}{\cite{CLCEL}:\\ $(n,k,0,\mu)$\\ AD set} & $n$ & $k$ & $k$ & $K$ & $K$ & $\frac{2n-2-k(k-1)}{2n}$ & \\ \hline
		\tabincell{c}{\cite{CLCEL2}:\\ $t$-$(n,k,\lambda)$\\ design with\\ $3t\geq2n$} & $n$ & $t$ & $t$ & $n$ & $n$ & $\frac{n-t}{n}$ & No \\ \hline
		\hline
		\tabincell{c}{Proposed:\\ $K/s\in\mathbb{N}$,\\ $r\in[K]$ and\\ $(r-1)/s\in\mathbb{N}$} & $K$ & $r$ & $s$ & \tabincell{c}{${K/s\choose \frac{r-1}{s}+1}\cdot$\\ $(r\!+\!s\!-\!1)$} & $K/s$ & $\frac{s(K-r+1)}{K(r+s-1)}$ & Yes \\ \hline
	\end{tabular}
	%
\end{table}

}

{ 

\subsection{Numerical Evaluations}
\label{subsec-Numerical Evaluations}
In this section, we present some numerical comparisons of the proposed scheme in Theorem~\ref{th-proposed scheme}, the new converse bound in Theorem~\ref{th-converse} compared to the well-known scheme in \cite{LMYA}, and the other grouping schemes in \cite{JQ} and \cite{WCJ}.
Recall that we focus on the case of $K/s\in\mathbb{N}$. In this case, there is no available structure for the schemes in \cite{JWZ,CWL,CLCEL,CLCEL2} which rely on several combinatorial design structures, thus we don't compare with them.

\begin{table}
	\center
	\vspace{-0.8cm}
	\caption{Communication load for the cascaded CDC with $K=100$ and $s=20$}\label{tab-K30s5}
	\resizebox{\linewidth}{!}{
		\begin{tabular}{ccccccccc}
			\toprule[1.5pt]
			& {$r=20$} & {$r=21$} & {$r=25$} & {$r=41$} & {$r=50$}  & {$r=61$} & {$r=81$}  \\ \hline
			Scheme in \cite{LMYA}  & $0.456$ & $0.440$ & $0.383$ & $0.227$ & $0.169$ & $0.114$ & $0.045$  \\
			Scheme in \cite{JQ}  &  $0.842$ & - & $0.625$& -& $0.204$ &  -& -\\
			Scheme in \cite{WCJ}  & $0.535$ & -& -& -& -& -& -\\
			Proposed scheme  & -& $\mathbf{0.400}$  & -& $\mathbf{\mathbf{0.200}}$ & -& $\mathbf{0.100}$ &  $\mathbf{\mathbf{0.040}}$  \\
			Converse bound  & - & $\mathbf{0.397}$ & - & $\mathbf{0.198}$ &-  & $\mathbf{0.098}$ & $\mathbf{0.038}$ \\
			\bottomrule[1.5pt]
	\end{tabular}}
\end{table}

\begin{figure}
	\centering
	\begin{subfigure}{0.45\textwidth}
		\includegraphics[width=3.3in]{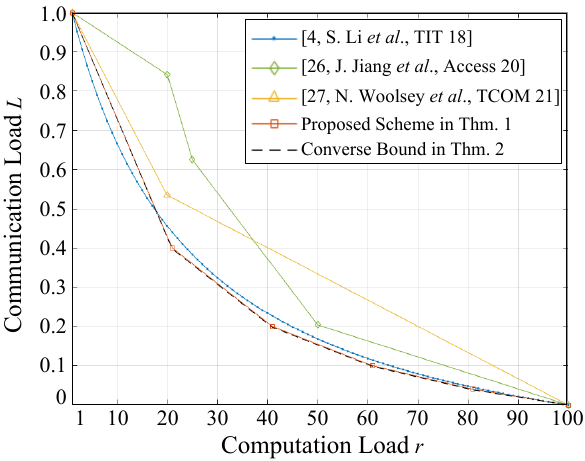}
		\caption{Comparison of the communication load $L$}
		\label{subfig-K30s5L}
	\end{subfigure} \ \ \ \ \ \ \
	\begin{subfigure}{0.45\textwidth}
		\includegraphics[width=3.3in]{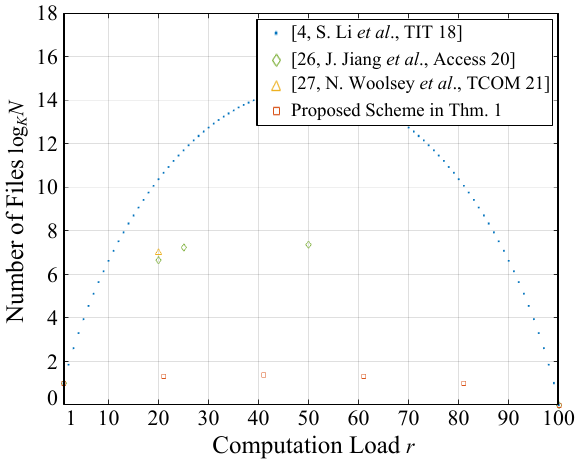}
		\caption{Comparison of the number of files $\log_{K}N$}
		\label{subfig-K30s5N}
	\end{subfigure}
	\caption{Comparisons of the communication load $L$ and number of files $\log_{K}N$ where $K=100$, $s=20$}\label{fig-K30s5}
\end{figure}

Here, we provide the numerical results for the case of $K=100$ and $s=20$, as illustrated in Table~\ref{tab-K30s5}\footnote{{ Considering the communication load $L(r,s)$, since all schemes satisfy that $L(1,s)=1$ and $L(K,s)=0$, in Table~\ref{tab-K30s5} the points $r=1$ and $r=100$ are omitted.}} and Fig.~\ref{fig-K30s5}.
Notice that, these schemes have their own applicable computation load $r$. From Table~\ref{tab-comp}, we have:
\begin{itemize}
	\item In \cite{LMYA}, $r\in[K]=[100]$. Due to space limitations, in Table~\ref{tab-K30s5} we only listed the points that need to be compared with other schemes.
	\item In \cite{JQ}, when $\frac{s}{r-1}(1-\frac{r}{K})<1$ and $K/r\in[N]$, the $r$ is available. Thus, we have $r\in\{20,25,50\}$.
	\item In \cite{WCJ}, since $r=s$ is required, only $r=20$ is available.
	\item In the proposed scheme, $r\in\{1,s+1,2s+1,\ldots,(K/s-1)s+1\}$ by Theorem~\ref{th-proposed scheme}. Thus, we have $r\in\{21,41,61,81\}$
\end{itemize}
For the other values of  $r$, it can be generalized by \cite[Section V-E]{LMYA}.
In Fig.~\ref{subfig-K30s5L}, the communication loads of the proposed scheme are clearly better than those of \cite{JQ,WCJ}, and some region of \cite{LMYA}.
\textit{Surprisingly, the deliberate data placement and output function assignment strategies proposed in this work allows us to achieve a lower communication load than \cite{LMYA} without the constraint $r=s=2$.}\footnote{{ To the best of our knowledge, 
in prior works with flexible in $K$, only \cite{WCJ} has achieved a communication load smaller than \cite{LMYA} under the specific constraint $r=s=2$. 
		This work is the first to do so without requiring $r=s=2$.}}
Moreover, the gap between the proposed scheme and the converse bound is exceedingly small.
In Fig.~\ref{subfig-K30s5N}, it shows that the proposed scheme has a significant advantage in the number of input files compared to all the results in \cite{LMYA}, \cite{JQ} and \cite{WCJ}.
Therefore, the proposed scheme has the potential to outperform other schemes supporting flexible $K$ in terms of both the communication load and the required number of files, and implements with one-shot delivery over the binary field.

}

\section{An Illustrative Example of the Proposed Coded Distributed Scheme}
\label{sec-example}
We consider the system with $K=6$ nodes that compute $Q=3$ output functions from $N=12$ input files, the computation load $r=3$, and each Reduce function is computed by $s=2$ nodes.  The main idea is that we divide the $K=6$ nodes  into $s=2$ groups as follows,
\begin{align}
	\mathcal{G}_1=\{1,2,3\}, \ \ \ \ \ \ \mathcal{G}_2=\{4,5,6\}, \label{eq-G1G2}
\end{align}
Then as Table \ref{tab-exam all} shows the stored input files, assigned output functions, and the delivered multicasting messages of the $6$ computing nodes, we deliberately design the data placement and output functions assignment by using the relationship between each group of nodes, such that only two-round transmissions are required during the Shuffle phase.
\begin{table}
\center
\caption{Files, functions, IVs, and multicasts of the $(K,r,s)=(6,3,2)$ cascaded CDC scheme}\label{tab-exam all}
\renewcommand\arraystretch{1}
\setlength{\tabcolsep}{0.6mm}{
\begin{tabular}{|c|c|c|c|c|c|c|}
\hline
nodes & files & computed IVs & functions & requested IVs & round I  & round II  \\ \hline
1 & $w_1$, $w_2$, $w_3$ & $v_{1,1}$, $v_{2,1}$, $v_{3,1}$, $v_{1,2}$, $v_{2,2}$, $v_{3,2}$, $v_{1,3}$, $v_{2,3}$, $v_{3,3}$ & $\phi_1$ & $v_{1,5}$, $v_{1,6}$, $v_{1,8}$ & $v_{1,2}^{(1)}\oplus v_{2,1}^{(1)}$ & $v_{2,4}\oplus v_{3,2}$ \\
 & $w_4$, $w_7$, $w_9$ &  $v_{1,4}$, $v_{2,4}$, $v_{3,4}$, $v_{1,7}$, $v_{2,7}$, $v_{3,7}$, $v_{1,9}$, $v_{2,9}$, $v_{3,9}$ & & $v_{1,10}$, $v_{1,11}$, $v_{1,12}$ & $v_{1,4}^{(1)}\oplus v_{3,3}^{(1)}$ &  \\ \hline
2 & $w_1$, $w_2$, $w_5$ & $v_{1,1}$, $v_{2,1}$, $v_{3,1}$, $v_{1,2}$, $v_{2,2}$, $v_{3,2}$, $v_{1,5}$, $v_{2,5}$, $v_{3,5}$ & $\phi_2$ & $v_{2,3}$, $v_{2,4}$, $v_{2,7}$ & $v_{1,2}^{(2)}\oplus v_{2,1}^{(2)}$ & $v_{1,6}\oplus v_{3,1}$ \\
& $w_6$, $w_8$, $w_{11}$ &  $v_{1,6}$, $v_{2,6}$, $v_{3,6}$, $v_{1,8}$, $v_{2,8}$, $v_{3,8}$, $v_{1,11}$, $v_{2,11}$, $v_{3,11}$ & & $v_{2,9}$, $v_{2,10}$, $v_{2,12}$ & $v_{2,6}^{(1)}\oplus v_{3,5}^{(1)}$ &  \\ \hline
3 & $w_3$, $w_4$, $w_5$ & $v_{1,3}$, $v_{2,3}$, $v_{3,3}$, $v_{1,4}$, $v_{2,4}$, $v_{3,4}$, $v_{1,5}$, $v_{2,5}$, $v_{3,5}$ & $\phi_3$ & $v_{3,1}$, $v_{3,2}$, $v_{3,7}$ & $v_{1,4}^{(2)}\oplus v_{3,3}^{(2)}$ & $v_{1,5}\oplus v_{2,3}$ \\
& $w_6$, $w_{10}$, $w_{12}$ &  $v_{1,6}$, $v_{2,6}$, $v_{3,6}$, $v_{1,10}$, $v_{2,10}$, $v_{3,10}$, $v_{1,12}$, $v_{2,12}$, $v_{3,12}$ & & $v_{3,8}$, $v_{3,9}$, $v_{3,11}$ & $v_{2,6}^{(2)}\oplus v_{3,5}^{(2)}$ &  \\ \hline
4 & $w_1$, $w_3$, $w_7$ & $v_{1,1}$, $v_{2,1}$, $v_{3,1}$, $v_{1,3}$, $v_{2,3}$, $v_{3,3}$, $v_{1,7}$, $v_{2,7}$, $v_{3,7}$ & $\phi_1$ & $v_{1,2}$, $v_{1,4}$, $v_{1,5}$ & $v_{1,8}^{(1)}\oplus v_{2,7}^{(1)}$ & $v_{2,10}\oplus v_{3,8}$ \\
& $w_8$, $w_9$, $w_{10}$ &  $v_{1,8}$, $v_{2,8}$, $v_{3,8}$, $v_{1,9}$, $v_{2,9}$, $v_{3,9}$, $v_{1,10}$, $v_{2,10}$, $v_{3,10}$ & & $v_{1,6}$, $v_{1,11}$, $v_{1,12}$ & $v_{1,10}^{(1)}\oplus v_{3,9}^{(1)}$ &  \\ \hline
5 & $w_2$, $w_5$, $w_7$ & $v_{1,2}$, $v_{2,2}$, $v_{3,2}$, $v_{1,5}$, $v_{2,5}$, $v_{3,5}$, $v_{1,7}$, $v_{2,7}$, $v_{3,7}$ & $\phi_2$ & $v_{2,1}$, $v_{2,3}$, $v_{2,4}$ & $v_{1,8}^{(2)}\oplus v_{2,7}^{(2)}$ & $v_{1,12}\oplus v_{3,7}$ \\
& $w_8$, $w_{11}$, $w_{12}$ &  $v_{1,8}$, $v_{2,8}$, $v_{3,8}$, $v_{1,11}$, $v_{2,11}$, $v_{3,11}$, $v_{1,12}$, $v_{2,12}$, $v_{3,12}$ & & $v_{2,6}$, $v_{2,9}$, $v_{2,10}$ & $v_{2,12}^{(1)}\oplus v_{3,11}^{(1)}$ &  \\ \hline
6 & $w_4$, $w_6$, $w_9$ & $v_{1,4}$, $v_{2,4}$, $v_{3,4}$, $v_{1,6}$, $v_{2,6}$, $v_{3,6}$, $v_{1,9}$, $v_{2,9}$, $v_{3,9}$ & $\phi_3$ & $v_{3,1}$, $v_{3,2}$, $v_{3,3}$ & $v_{1,10}^{(2)}\oplus v_{3,9}^{(2)}$ & $v_{1,11}\oplus v_{2,9}$ \\
& $w_{10}$, $w_{11}$, $w_{12}$ &  $v_{1,10}$, $v_{2,10}$, $v_{3,10}$, $v_{1,11}$, $v_{2,11}$, $v_{3,11}$, $v_{1,12}$, $v_{2,12}$, $v_{3,12}$ & & $v_{3,5}$, $v_{3,7}$, $v_{3,8}$ & $v_{2,12}^{(2)}\oplus v_{3,11}^{(2)}$ &  \\ \hline
\end{tabular}}
\end{table}
Now let us introduce 1) the data placement; 2) the output functions assignment; and 3) the
Shuffle strategy of two rounds in the following.

\textit{1) Data Placement}: Recall that $\mathcal{D}_n$, $n\in[N]$ presents the set of nodes storing the input file $w_n$. For each $n\in[12]$,  let $|\mathcal{D}_n|=r=3$, i.e., each file is stored by $3$ nodes.
To determine which nodes a file is stored by, we first select $1$ group of nodes, and choose $(r-1)/s+1=2$ nodes of it; then for the nodes in another group, we can choose any $(r-1)/s=1$ node with the same position as the previously selected nodes.
Hence, there are $N=s{K/Q\choose (r-1)/s+1}{(r-1)/s+1\choose (r-1)/s}=12$ different input files.
For instance, $w_1$ is stored by nodes ${1,2,4}$, where nodes ${1,2}$ are the first and second elements belonging to  $\mathcal{G}_1$, thus we can choose node ${4}$ which is the first element  belonging to $\mathcal{G}_2$, then we have $\mathcal{D}_1=\{1,2,4\}$.
Similarly, all the input files are stored as follows,
\begin{align}
&\mathcal{D}_1=\{1,2,4\}, \ \
\mathcal{D}_2=\{1,2,5\}, \ \
\mathcal{D}_3=\{1,3,4\},  \nonumber\\
&\mathcal{D}_4=\{1,3,6\}, \ \
\mathcal{D}_5=\{2,3,5\},  \ \
\mathcal{D}_6=\{2,3,6\}, \label{eq-A1-A12}\\
&\mathcal{D}_7=\{1,4,5\},  \ \
\mathcal{D}_8=\{2,4,5\}, \ \
\mathcal{D}_9=\{1,4,6\},\nonumber\\    &\mathcal{D}_{10}=\{3,4,6\}, \ \
\mathcal{D}_{11}=\{2,5,6\},  \ \
\mathcal{D}_{12}=\{3,5,6\}.\nonumber
\end{align}
By \eqref{eq-model Mk}, we obtain the stored input files for each node as shown in the second column of Table~\ref{tab-exam all}.

\textit{2) Output Functions Assignment}:
Recall that $\mathcal{A}_q$, $q\in[Q]$ represents the set of nodes  that are assigned with the output function $\phi_q$. Since $s=2$, i.e., each function is assigned to $2$ nodes, we have $|\mathcal{A}_q|=2$ where $q\in[3]$.
For an output function,  assign it to $s$ nodes, which are distributed in $s$ groups with the same position. Hence, there are $Q=K/s=3$ different output functions.
For instance, $\phi_1$ is assigned to nodes $1,4$, both of which are the first element of $\mathcal{G}_1$ and $\mathcal{G}_2$. Thus we have $\mathcal{A}_1=\{1,4\}$.
Similarly, all the output functions are assigned as follows,
\begin{align}
\mathcal{A}_1=\{1,4\}, \ \ \ \ \ \ \mathcal{A}_2=\{2,5\}, \ \ \ \ \ \ \mathcal{A}_3=\{3,6\}. \label{eq-B1-b3}
\end{align}
By \eqref{eq-model Wk}, we obtain the assigned output functions for each node as shown in the fourth column of Table~\ref{tab-exam all}.

\textit{3) Shuffle Strategy of Two Rounds}:
In the Map phase, each node can compute all the IVs from its stored files, as shown in the third column of Table~\ref{tab-exam all}.
For instance, node 1 can compute the IVs $\{v_{q,n} : q\in[3],w_n\in\mathcal{M}_1\}$.
Thus there are  $N\times Q=12\times3$ different IVs $\{v_{q,n}:q\in[3],n\in[12]\}$ computed by all the $6$ nodes.
We classify these IVs into 3 categories based on the number of nodes requesting them during the Shuffle phase.
\begin{itemize}
\item { IVs not requested by any node: The following IVs are not requested by any node,}
\begin{align}
v_{1,1}, \ v_{1,3}, \ v_{1,7}, \ v_{1,9}, \
v_{2,2}, \ v_{2,5}, \ v_{2,8}, \ v_{2,11}, \
v_{3,4}, \ v_{3,6}, \ v_{3,10}, \ v_{3,12}, \label{eq-IVs 0 node}
\end{align}
which means these IVs do not need to be transmitted during the Shuffle phase.
For instance, $v_{1,1}$ is needed by nodes $1$ and $4$ since they are assigned function $\phi_1$, while these nodes are placed the file $w_1$ such that they can compute $v_{1,1}$ in the Map phase. Thus  $v_{1,1}$ is requested by $0$ node.

\item IVs requested by $s-1=1$ node: The following IVs are requested by $1$ node,
\begin{align}
v_{1,2}, \ v_{1,4}, \ v_{1,8}, \ v_{1,10}, \
v_{2,1}, \ v_{2,6}, \ v_{2,7}, \ v_{2,12}, \
v_{3,3}, \ v_{3,5}, \ v_{3,9}, \ v_{3,11}. \label{eq-IVs 1 node}
\end{align}
For instance, $v_{1,2}$ is needed by nodes $1$ and $4$, and node $1$ can compute $v_{1,2}$ since it stores the file $w_2$, thus $v_{1,2}$ is only requested by $1$ node (node $4$).

\item IVs requested by $s=2$ nodes: The following IVs are requested by $2$ nodes,
\begin{align}
v_{1,5}, \ v_{1,6}, \ v_{1,11}, \ v_{1,12}, \
v_{2,3}, \ v_{2,4}, \ v_{2,9}, \ v_{2,10}, \
v_{3,1}, \ v_{3,2}, \ v_{3,7}, \ v_{3,8}. \label{eq-IVs 2 nodes}
\end{align}
For instance, $v_{1,5}$ is requested by $2$ nodes (nodes $1$ and $4$), since neither node are placed the file $w_5$ to compute $v_{1,5}$.
\end{itemize}

There are two rounds of transmission in the Shuffle phase, where the multicast messages by each node are shown in the last columns of Table~\ref{tab-exam all}, respectively.
In round $1$, we consider the IVs in \eqref{eq-IVs 1 node} which are requested by $1$ node. Each IV is split into $(r-1)/s+1=2$ disjoint equal-sized packets.
For instance, $v_{1,2}$ is split into $v_{1,2}^{(1)}$ and $v_{1,2}^{2}$, which are transmitted by nodes $1$ and $2$, respectively.
Each node transmits $2$ bit-wise XOR of two available packets, where each XOR is multicasted to the other $2$ nodes. Furthermore, each node recovers a requested IV by $2$ XOR from $2$ different nodes.
For instance, node $1$ multicasts $v_{1,2}^{(1)}\bigoplus v_{2,1}^{(1)}$ to nodes $4$ and $5$, where $v_{1,2}^{(1)}$ is requested by node $4$ and available at node $5$, and the opposite is true for $v_{2,1}^{(1)}$, thus they can decode $v_{1,2}^{(1)}$ and $v_{2,1}^{(1)}$, respectively. Moreover, nodes $4$ and $5$ can receive the XOR $v_{1,2}^{(2)}\bigoplus v_{2,1}^{(2)}$ from node $2$, such that they can decode $v_{1,2}^{(2)}$ and $v_{2,1}^{(2)}$, respectively.
Hence, nodes $4$ and $5$ recover $v_{1,2}$ and $v_{2,1}$, respectively.

In round $2$, we consider the IVs in \eqref{eq-IVs 2 nodes} which are requested by $2$ nodes.
Each node multicasts an XOR of two locally computed IVs to the other $4$ nodes.
For instance, node $1$ multicasts $v_{2,4}\bigoplus v_{3,2}$ to nodes $2$, $3$, $5$ and $6$. Since nodes $2$, $5$ (the second element of $\mathcal{G}_1$ and $\mathcal{G}_2$) know $v_{3,2}$ locally, thus they can decode the requested $v_{2,4}$. Similarly, nodes $3$ and $6$ (the third element of $\mathcal{G}_1$ and $\mathcal{G}_2$)  can decode the requested $v_{3,2}$.

In the Shuffle strategy of two rounds, there are $12$ messages of length $T/2$ bits ($T$ is the size of a single IV), and $6$  messages of length $T$ bits. By \eqref{eq-def of L}, the communication load is,
\begin{align}
L=\frac{1}{QNT}\left(12\cdot\frac{T}{2}+6T\right)=\frac{12}{36}=0.333,
\end{align}
which coincides with \eqref{eq-Load}.

\section{Proof of Theorem~\ref{th-proposed scheme}}
\label{sec-proof}
In this section, we formally prove  Theorem~\ref{th-proposed scheme} by presenting the cascaded CDC scheme.
We first consider the specific number of input files $N_1$ and output functions $Q_1$, where $N_1=(r+s-1){K/s\choose (r-1)/s+1}$ and $Q_1=K/s$ (given by \eqref{eq-N} and \eqref{eq-Q}, respectively). Then, for the general case of $N/N_1\in\mathbb{N}$ and $Q/Q_1\in\mathbb{N}$, the $N$ input files are evenly partitioned into $N_1$ disjoint batches of size $N/N_1$, each corresponding to a unique subset of $r$ nodes; and the $Q$ output functions are evenly partitioned into $Q_1$ disjoint batches of size $Q/Q_1$, each corresponding to a unique subset of $s$ nodes. To simplify the notations, let $\Lambda=K/s$, $t=(r-1)/s$.
	
Firstly, we divide the $K$ nodes into $s$ groups with equal length. For $i\in[s]$, the $i$th group of nodes is denoted by
\begin{align}
\mathcal{G}_i=\left\{(i-1)\Lambda+1, (i-1)\Lambda+2,\ldots, i\Lambda\right\},\label{eq-group}
\end{align}
where $\Lambda=K/s$.
Thus $|\mathcal{G}_i|=\Lambda$, i.e., each group contains $\Lambda$ nodes.
Note that, when $K=6$, $r=3$, $s=2$, the grouping denoted by \eqref{eq-group} is illustrated in \eqref{eq-G1G2}.
Based on the grouping method, the scheme is described by the following three parts: 1) data placement; 2) output functions assignment; and 3) Shuffle strategy of two rounds.

\subsection{Data Placement}
\label{subsec-input store}
For simplicity of expression, for each $n\in[N_1]$, the subscript $n$ of input file $w_n$ is denoted by a tuple $(i,\mathcal{C},\mathcal{T})\in\mathcal{N}$, and formulated by
\begin{align}
\mathcal{N}=\left\{(i,\mathcal{C},\mathcal{T}) \ : \ \ i\in[s], \ \mathcal{C}\in{[\Lambda]\choose t+1}, \ \mathcal{T}\in{\mathcal{C}\choose t}\right\}, \label{eq-n index}
\end{align}
where $t=(r-1)/s$.
Furthermore,
\begin{align}
	N_1=|\mathcal{N}|=s{\Lambda\choose t+1}{t+1\choose t}=s(t+1){\Lambda\choose t+1}=(r+s-1){K/s\choose (r-1)/s+1}, \label{eq-N}
\end{align}
which gives the number of input files that coincide with Theorem~\ref{th-proposed scheme}.

The file $w_{(i,\mathcal{C},\mathcal{T})}$ is stored by nodes in\footnote{{ Recall that for any non-negative set $\mathcal{V}$ and  $\mathcal{J}$, $\mathcal{V}[\mathcal{J}]:=\{\mathcal{V}_j:j\in\mathcal{J}\}$. 
In other words, $\mathcal{V}[\mathcal{J}]$ represents a subset of $\mathcal{V}$ consisting  of elements indexed by $\mathcal{J}$, where the elements of each set are listed in increasing order.  } }
\begin{align}
\mathcal{D}_{(i,\mathcal{C},\mathcal{T})}=\left\{\mathcal{G}_i[\mathcal{C}]\bigcup_{j\in[s]\setminus\{i\}}\mathcal{G}_j[\mathcal{T}] : i\in[s],  \mathcal{C}\in{[\Lambda]\choose t+1},  \mathcal{T}\in{\mathcal{C}\choose t}\right\}. \label{eq-file store}
\end{align}
Thus each file is stored by $|\mathcal{D}_{(i,\mathcal{C},\mathcal{T})}|=(t+1)+(s-1)t=r$ computing nodes.

Then, for each node $k\in[K]$, it stores the input files as follows,
\begin{align}
	\mathcal{M}_k=\{w_{(i,\mathcal{C},\mathcal{T})} : k\in\mathcal{D}_{(i,\mathcal{C},\mathcal{T})}\}.
\end{align}
We can check that the computation load constraint is satisfied, since
\begin{align}	\frac{\sum_{k\in[K]}|\mathcal{M}_k|}{N}
=&\frac{K\cdot\left({\Lambda-1\choose t}(t+1)+(s-1){\Lambda-1\choose t}{t\choose t-1}\right)}{s(t+1){\Lambda\choose t+1}}\nonumber\\
=&\frac{s\Lambda\cdot{\Lambda-1\choose t}(st+1)}{s(t+1){\Lambda\choose t+1}}=st+1=r.
\end{align}

\begin{example}
When $K=6$, $r=3$, $s=2$, and $N/N_1=1$, from \eqref{eq-n index} and \eqref{eq-file store}, we have the input files stored as follows.
\begin{align}
&\mathcal{D}_{(1,\{1,2\},\{1\})}
=\{\mathcal{G}_1[\{1,2\}]\cup\mathcal{G}_2[\{1\}]\}=\{1,2,4\},
\nonumber\\
&\mathcal{D}_{(1,\{1,2\},\{2\})}
=\{\mathcal{G}_1[\{1,2\}]\cup\mathcal{G}_2[\{2\}]\}=\{1,2,5\}, \nonumber\\
&\mathcal{D}_{(1,\{1,3\},\{1\})}
=\{\mathcal{G}_1[\{1,3\}]\cup\mathcal{G}_2[\{1\}]\}=\{1,3,4\},
\nonumber\\
&\mathcal{D}_{(1,\{1,3\},\{3\})}
=\{\mathcal{G}_1[\{1,3\}]\cup\mathcal{G}_2[\{3\}]\}=\{1,3,6\}, \nonumber\\
&\mathcal{D}_{(1,\{2,3\},\{2\})}
=\{\mathcal{G}_1[\{2,3\}]\cup\mathcal{G}_2[\{2\}]\}=\{2,3,5\},
\nonumber\\
&\mathcal{D}_{(1,\{2,3\},\{3\})}
=\{\mathcal{G}_1[\{2,3\}]\cup\mathcal{G}_2[\{3\}]\}=\{2,3,6\}, \nonumber\\
&\mathcal{D}_{(2,\{1,2\},\{1\})}
=\{\mathcal{G}_2[\{1,2\}]\cup\mathcal{G}_1[\{1\}]\}=\{1,4,5\},
\nonumber\\
&\mathcal{D}_{(2,\{1,2\},\{2\})}
=\{\mathcal{G}_2[\{1,2\}]\cup\mathcal{G}_1[\{2\}]\}=\{2,4,5\}, \nonumber\\
&\mathcal{D}_{(2,\{1,3\},\{1\})}
=\{\mathcal{G}_2[\{1,3\}]\cup\mathcal{G}_1[\{1\}]\}=\{1,4,6\},
\nonumber\\
&\mathcal{D}_{(2,\{1,3\},\{3\})}
=\{\mathcal{G}_2[\{1,3\}]\cup\mathcal{G}_1[\{3\}]\}=\{3,4,6\}, \nonumber\\
&\mathcal{D}_{(2,\{2,3\},\{2\})}
=\{\mathcal{G}_2[\{2,3\}]\cup\mathcal{G}_1[\{2\}]\}=\{2,5,6\},
\nonumber\\
&\mathcal{D}_{(2,\{2,3\},\{3\})}
=\{\mathcal{G}_2[\{2,3\}]\cup\mathcal{G}_1[\{3\}]\}=\{3,5,6\}. \label{eq-D exam}
\end{align}

By replacing the subscript tuples $(i,\mathcal{C},\mathcal{T})$ of \eqref{eq-D exam} into integers $n\in[N]=[12]$ according to the one-to-one mapping $\varphi$ in Table~\ref{tab-mapping}, we have the results as illustrated in \eqref{eq-A1-A12}. Thus, the stored input files for each node $k\in[6]$ coincide with the second column of Table~\ref{tab-exam all}.
\begin{table}
\center
\caption{The mapping $\varphi$ }\label{tab-mapping}
\begin{tabular}{|c|c|c|c|c|c|c|}
\hline
$(i,\mathcal{C},\mathcal{T})$ & $(1,\{1,2\},\{1\})$ & $(1,\{1,2\},\{2\})$ & $(1,\{1,3\},\{1\})$ & $(1,\{1,3\},\{3\})$ & $(1,\{2,3\},\{2\})$ & $(1,\{2,3\},\{3\})$ \\ \hline
$n$ & $1$ & $2$ & $3$ & $4$ & $5$ & $6$  \\ \hline
$(i,\mathcal{C},\mathcal{T})$ & $(2,\{1,2\},\{1\})$ & $(2,\{1,2\},\{2\})$ & $(2,\{1,3\},\{1\})$ & $(2,\{1,3\},\{3\})$ & $(2,\{2,3\},\{2\})$ & $(2,\{2,3\},\{3\})$ \\ \hline
$n$ & $7$ & $8$ & $9$ & $10$ & $11$ & $12$  \\ \hline
\end{tabular}
\end{table}
\end{example}

\subsection{Output Functions Assignment}
\label{subsec-output assign}
For $q\in[Q_1]$,  define the output function assignment as follows,
\begin{align}
\mathcal{A}_q=\{\mathcal{G}_i[q] : i\in[s]\}, \label{eq-out assign}
\end{align}
{ i.e., the output function $\phi_q$ is assigned to the nodes in $\mathcal{A}_q$}. Thus there are 
\begin{align}
	Q_1=\Lambda=\frac{K}{s} \label{eq-Q}
\end{align}
different integers $q$, which gives the number of output functions  that coincide with Theorem~\ref{th-proposed scheme}.
Furthermore,  each function is assigned to $|\mathcal{A}_q|=s$ computing nodes.

Then, for each node $k\in[K]$, it is assigned the output functions as follows,
\begin{align}
\mathcal{W}_k=\{\phi_q:k\in\mathcal{A}_q\}.
\end{align}

Note that, when $K=6$, $r=3$, $s=2$, and $Q/Q_1=1$, the output function assignment denoted by $\eqref{eq-out assign}$ is illustrated in \eqref{eq-B1-b3}.
Thus, the assigned output functions for each node $k\in[6]$ coincide with the fourth column of Table~\ref{tab-exam all}.

\subsection{Two-rounds Shuffle Phase and Reduce Phase}
\label{subsec-shuffle}
There are $N_1\times Q_1$ different IVs overall, $\{v_{q,(i,\mathcal{C},\mathcal{T})}: q\in[Q_1],(i,\mathcal{C},\mathcal{T})\in\mathcal{N}\}$, where $v_{q,(i,\mathcal{C},\mathcal{T})}$ is locally mapped by the nodes in $\mathcal{D}_{(i,\mathcal{C},\mathcal{T})}$ and requested by the nodes in $\mathcal{A}_q$ during the Shuffle phase.
For these IVs, based on the number of nodes requesting them during the Shuffle phase, they are classified into 3 categories: { \textit{Type I} are not requested by any node}; \textit{Type II} requested by $s-1$ nodes; and \textit{Type III} requested by $s$ nodes.
Thus there are two rounds of transmission during the Shuffle phase, to deal with  Type II and Type III IVs, respectively.

\begin{itemize}
\item Type I IVs: { The IVs not requested by any node}, which are formulated by
\begin{align}
\mathcal{V}_{\text{I}}&=\left\{v_{q,(i,\mathcal{C},\mathcal{T})}:\mathcal{A}_q\subseteq\mathcal{D}_{(i,\mathcal{C},\mathcal{T})}\right\} \nonumber \\ &=\left\{v_{q,(i,\mathcal{C},\mathcal{T})}:\forall (i,\mathcal{C,\mathcal{T}})\in\mathcal{N}, q\in\mathcal{T}\right\}\label{eq-IVs I}.
\end{align}	
Thus the number of IVs belonging to Type I is
\begin{align}
F_0=|\mathcal{V}_{\text{I}}|=N_1t. \label{eq-F0}
\end{align}

{ \begin{remark}
		\label{rem-s=1 F0'}
		When $s=1$, there are $N_1'r$ distinct IVs belonged to Type I.
		In this case, we have $\Lambda=K/s=K$ and $t=(r-1)/s=r-1$. Then, \eqref{eq-n index} is reduced to $\mathcal{N}'=\left\{(1,\mathcal{C},\mathcal{T}):\mathcal{C}\in{[K]\choose r},\mathcal{T}\in{\mathcal{C}\choose r-1}\right\}$, and $N_1'=|\mathcal{N}'|=r{K\choose r}$;
		\eqref{eq-out assign} is reduced to $\mathcal{A}'_{q'}=\{\mathcal{G}_1[q']\}=\{q'\}$, and $Q'=K$. 
		Hence, there are $F_0'=|\mathcal{V}_{\text{I}}'|=N_1'r$ distinct IVs belonged to Type I, since  $\mathcal{V}_{\text{I}}'=\{v_{q',(1,\mathcal{C},\mathcal{T})}:\forall{(1,\mathcal{C},\mathcal{T})\in\mathcal{N}'},q'\in\mathcal{T}\}$.
\end{remark}}

\vspace{1mm}
	
\item Type II IVs: The IVs requested by $s-1$ nodes, which are formulated by
\begin{align}
\mathcal{V}_{\text{II}}&=\left\{v_{q,(i,\mathcal{C},\mathcal{T})}:|\mathcal{A}_q\cap\mathcal{D}_{(i,\mathcal{C},\mathcal{T})}|=1\right\} \nonumber \\
&=\{v_{q,(i,\mathcal{C},\mathcal{T})} : \forall (i,\mathcal{C},\mathcal{T})\in\mathcal{N},q=\mathcal{C}\setminus\mathcal{T}\} \label{eq-IVs II}.
\end{align}
Thus the number of IVs belonging to Type II is
\begin{align}
F_1=|\mathcal{V}_{\text{II}}|=N_1. \label{eq-F1}
\end{align}

\begin{remark}
\label{rem-s=1 F1'}
When $s=1$, i.e., $s-1=0$, there is $F_1'=0$ IV  belonging to Type II. Thus there is only one round (the round 2) transmission during the Shuffle phase.
\end{remark}

\item Type III IVs: The set of IVs requested by $s$ nodes, which is formulated by
\begin{align}
\mathcal{V}_{\text{III}}&=\left\{v_{q,(i,\mathcal{C},\mathcal{T})}:\mathcal{A}_q\cap\mathcal{D}_{(i,\mathcal{C},\mathcal{T})}=\emptyset\right\} \nonumber \\
&=\{v_{q,(i,\mathcal{C},\mathcal{T})}: \forall (i,\mathcal{C},\mathcal{T})\in\mathcal{N},q\in[\Lambda]\setminus\mathcal{C}\}\label{eq-IVs III}.
\end{align}
Thus the number of IVs belonging to Type III is
\begin{align}
F_2=|\mathcal{V}_{\text{III}}|=N_1(\Lambda-(t+1)). \label{eq-F2}
\end{align}
\begin{remark}
When $K=r+s-1$, i.e., $\Lambda=t+1$, we have $F_2=0$, thus there is only one round (the round 1) transmission during the Shuffle phase.
\end{remark}	
\begin{remark}
\label{rem-s=1 F2'}
	When $s=1$, there are $F_2'=|\mathcal{V}_{\text{III}}'|=N_1'(K-r)$ distinct IVs belonging to Type III, since $\mathcal{V}_{\text{III}}'=\{v_{q',(1,\mathcal{C},\mathcal{T})}:\forall(1,\mathcal{C},\mathcal{T})\in\mathcal{N}',q'\in[K]\setminus\mathcal{C})\}$.
\end{remark}
\end{itemize}

We can check that, $F_0+F_1+F_2=N_1t+N_1+N_1(\Lambda-(t+1))=N_1\Lambda=N_1\times Q_1$, { i.e., the all IVs are precisely classified into three types}.
Furthermore, when $s=1$, from Remarks \ref{rem-s=1 F0'}, \ref{rem-s=1 F1'} and \ref{rem-s=1 F2'}, we have that $F_0'+F_1'+F_2'=N_1'r+0+N_1'(K-r)=N_1'K=N_1'\times Q_1'$.

\begin{example}
Let us return to the example in Section~\ref{sec-example} with $K=6$, $r=3$, and $s=2$.
From \eqref{eq-IVs I}, \eqref{eq-D exam}, and \eqref{eq-B1-b3}, we have the type I IVs as follows.
\begin{align}
v_{1,(1,\{1,2\},\{1\})} \ \ \ \ \ \ \text{since} \ \mathcal{A}_1=\{1,4\}\in\mathcal{D}_{(1,\{1,2\},\{1\})}=\{1,2,4\}, \nonumber \\
v_{1,(1,\{1,3\},\{1\})} \ \ \ \ \ \ \text{since} \ \mathcal{A}_1=\{1,4\}\in\mathcal{D}_{(1,\{1,3\},\{1\})}=\{1,3,4\}, \nonumber \\
v_{1,(2,\{1,2\},\{1\})} \ \ \ \ \ \ \text{since} \ \mathcal{A}_1=\{1,4\}\in\mathcal{D}_{(2,\{1,2\},\{1\})}=\{1,4,5\}, \nonumber \\
v_{1,(2,\{1,3\},\{1\})} \ \ \ \ \ \ \text{since} \ \mathcal{A}_1=\{1,4\}\in\mathcal{D}_{(2,\{1,3\},\{1\})}=\{1,4,6\}, \nonumber \\
v_{2,(1,\{1,2\},\{2\})} \ \ \ \ \ \ \text{since} \ \mathcal{A}_2=\{2,5\}\in\mathcal{D}_{(1,\{1,2\},\{2\})}=\{1,2,5\}, \nonumber \\
v_{2,(1,\{2,3\},\{2\})} \ \ \ \ \ \ \text{since} \ \mathcal{A}_2=\{2,5\}\in\mathcal{D}_{(1,\{2,3\},\{2\})}=\{2,3,5\}, \nonumber \\
v_{2,(2,\{1,2\},\{2\})} \ \ \ \ \ \ \text{since} \ \mathcal{A}_2=\{2,5\}\in\mathcal{D}_{(2,\{1,2\},\{2\})}=\{2,4,5\}, \nonumber \\
v_{2,(2,\{2,3\},\{2\})} \ \ \ \ \ \ \text{since} \ \mathcal{A}_2=\{2,5\}\in\mathcal{D}_{(2,\{2,3\},\{2\})}=\{2,5,6\}, \nonumber \\
v_{3,(1,\{1,3\},\{3\})} \ \ \ \ \ \ \text{since} \ \mathcal{A}_3=\{3,6\}\in\mathcal{D}_{(1,\{1,3\},\{3\})}=\{1,3,6\}, \nonumber \\
v_{3,(1,\{2,3\},\{3\})} \ \ \ \ \ \ \text{since} \ \mathcal{A}_3=\{3,6\}\in\mathcal{D}_{(1,\{2,3\},\{3\})}=\{2,3,6\}, \nonumber \\
v_{3,(2,\{1,3\},\{3\})} \ \ \ \ \ \ \text{since} \ \mathcal{A}_3=\{3,6\}\in\mathcal{D}_{(2,\{1,3\},\{3\})}=\{3,4,6\}, \nonumber \\
v_{3,(2,\{2,3\},\{3\})} \ \ \ \ \ \ \text{since} \ \mathcal{A}_3=\{3,6\}\in\mathcal{D}_{(2,\{2,3\},\{3\})}=\{3,5,6\}. \label{eq-IV 0 exam}
\end{align}
By replacing the subscript tuples $(i,\mathcal{C},\mathcal{T})$ of \eqref{eq-IV 0 exam} into integers $n\in[N]=[12]$ according to the one-to-one mapping $\varphi$ in Table~\ref{tab-mapping}, we have the results as illustrated in \eqref{eq-IVs 0 node}.

From \eqref{eq-IVs II}, \eqref{eq-D exam}, and \eqref{eq-B1-b3}, we have the type II IVs as follows.
\begin{align}
v_{1,(1,\{1,2\},\{2\})} \ \ \ \ \ \ \text{since} \ |\mathcal{A}_1\cap\mathcal{D}_{(1,\{1,2\},\{2\})}|=|\{1,4\}\cap\{1,2,5\}|=1, \nonumber \\
v_{1,(1,\{1,3\},\{3\})} \ \ \ \ \ \ \text{since} \ |\mathcal{A}_1\cap\mathcal{D}_{(1,\{1,3\},\{3\})}|=|\{1,4\}\cap\{1,3,6\}|=1, \nonumber \\
v_{1,(2,\{1,2\},\{2\})} \ \ \ \ \ \ \text{since} \ |\mathcal{A}_1\cap\mathcal{D}_{(2,\{1,2\},\{2\})}|=|\{1,4\}\cap\{2,4,5\}|=1, \nonumber \\
v_{1,(2,\{1,3\},\{3\})} \ \ \ \ \ \ \text{since} \ |\mathcal{A}_1\cap\mathcal{D}_{(2,\{1,3\},\{3\})}|=|\{1,4\}\cap\{3,4,6\}|=1, \nonumber \\
v_{2,(1,\{1,2\},\{1\})} \ \ \ \ \ \ \text{since} \ |\mathcal{A}_2\cap\mathcal{D}_{(1,\{1,2\},\{1\})}|=|\{2,5\}\cap\{1,2,4\}|=1, \nonumber \\
v_{2,(1,\{2,3\},\{3\})} \ \ \ \ \ \ \text{since} \ |\mathcal{A}_2\cap\mathcal{D}_{(1,\{2,3\},\{3\})}|=|\{2,5\}\cap\{2,3,6\}|=1, \nonumber \\
v_{2,(2,\{1,2\},\{1\})} \ \ \ \ \ \ \text{since} \ |\mathcal{A}_2\cap\mathcal{D}_{(2,\{1,2\},\{1\})}|=|\{2,5\}\cap\{1,4,5\}|=1, \nonumber \\
v_{2,(2,\{2,3\},\{3\})} \ \ \ \ \ \ \text{since} \ |\mathcal{A}_2\cap\mathcal{D}_{(2,\{2,3\},\{3\})}|=|\{2,5\}\cap\{3,5,6\}|=1, \nonumber \\
v_{3,(1,\{1,3\},\{1\})} \ \ \ \ \ \ \text{since} \ |\mathcal{A}_3\cap\mathcal{D}_{(1,\{1,3\},\{1\})}|=|\{3,6\}\cap\{1,3,4\}|=1, \nonumber \\
v_{3,(1,\{2,3\},\{2\})} \ \ \ \ \ \ \text{since} \ |\mathcal{A}_3\cap\mathcal{D}_{(1,\{2,3\},\{2\})}|=|\{3,6\}\cap\{2,3,5\}|=1, \nonumber \\
v_{3,(2,\{1,3\},\{1\})} \ \ \ \ \ \ \text{since} \ |\mathcal{A}_3\cap\mathcal{D}_{(2,\{1,3\},\{1\})}|=|\{3,6\}\cap\{1,4,6\}|=1, \nonumber \\
v_{3,(2,\{2,3\},\{2\})} \ \ \ \ \ \ \text{since} \ |\mathcal{A}_3\cap\mathcal{D}_{(2,\{2,3\},\{2\})}|=|\{3,6\}\cap\{2,5,6\}|=1. \label{eq-IV 1 exam}
\end{align}
By replacing the subscript tuples $(i,\mathcal{C},\mathcal{T})$ of \eqref{eq-IV 1 exam} into integers $n\in[N]=[12]$ according to the one-to-one mapping $\varphi$ in Table~\ref{tab-mapping}, we have the results as illustrated in \eqref{eq-IVs 1 node}.

From \eqref{eq-IVs III}, \eqref{eq-D exam}, and \eqref{eq-B1-b3}, we have the type III IVs as follows.
\begin{align}
v_{1,(1,\{2,3\},\{2\})} \ \ \ \ \ \ \text{since} \ \mathcal{A}_1\cap\mathcal{D}_{(1,\{2,3\},\{2\})}=\{1,4\}\cap\{2,3,5\}=\emptyset, \nonumber \\
v_{1,(1,\{2,3\},\{3\})} \ \ \ \ \ \ \text{since} \ \mathcal{A}_1\cap\mathcal{D}_{(1,\{2,3\},\{3\})}=\{1,4\}\cap\{2,3,6\}=\emptyset, \nonumber \\
v_{1,(2,\{2,3\},\{2\})} \ \ \ \ \ \ \text{since} \ \mathcal{A}_1\cap\mathcal{D}_{(2,\{2,3\},\{2\})}=\{1,4\}\cap\{2,5,6\}=\emptyset, \nonumber \\
v_{1,(2,\{2,3\},\{3\})} \ \ \ \ \ \ \text{since} \ \mathcal{A}_1\cap\mathcal{D}_{(2,\{2,3\},\{3\})}=\{1,4\}\cap\{3,5,6\}=\emptyset, \nonumber \\
v_{2,(1,\{1,3\},\{1\})} \ \ \ \ \ \ \text{since} \ \mathcal{A}_2\cap\mathcal{D}_{(1,\{1,3\},\{1\})}=\{2,5\}\cap\{1,3,4\}=\emptyset, \nonumber \\
v_{2,(1,\{1,3\},\{3\})} \ \ \ \ \ \ \text{since} \ \mathcal{A}_2\cap\mathcal{D}_{(1,\{1,3\},\{3\})}=\{2,5\}\cap\{1,3,6\}=\emptyset, \nonumber \\
v_{2,(2,\{1,3\},\{1\})} \ \ \ \ \ \ \text{since} \ \mathcal{A}_2\cap\mathcal{D}_{(2,\{1,3\},\{1\})}=\{2,5\}\cap\{1,4,6\}=\emptyset, \nonumber \\
v_{2,(2,\{1,3\},\{3\})} \ \ \ \ \ \ \text{since} \ \mathcal{A}_2\cap\mathcal{D}_{(2,\{1,3\},\{3\})}=\{2,5\}\cap\{3,4,6\}=\emptyset, \nonumber \\
v_{3,(1,\{1,2\},\{1\})} \ \ \ \ \ \ \text{since} \ \mathcal{A}_3\cap\mathcal{D}_{(1,\{1,2\},\{1\})}=\{3,6\}\cap\{1,2,4\}=\emptyset, \nonumber \\
v_{3,(1,\{1,2\},\{2\})} \ \ \ \ \ \ \text{since} \ \mathcal{A}_3\cap\mathcal{D}_{(1,\{1,2\},\{2\})}=\{3,6\}\cap\{1,2,5\}=\emptyset, \nonumber \\
v_{3,(2,\{1,2\},\{1\})} \ \ \ \ \ \ \text{since} \ \mathcal{A}_3\cap\mathcal{D}_{(2,\{1,2\},\{1\})}=\{3,6\}\cap\{1,4,5\}=\emptyset, \nonumber \\
v_{3,(2,\{1,2\},\{2\})} \ \ \ \ \ \ \text{since} \ \mathcal{A}_3\cap\mathcal{D}_{(2,\{1,2\},\{2\})}=\{3,6\}\cap\{2,4,5\}=\emptyset. \label{eq-IV 2 exam}
\end{align}
By replacing the subscript tuples $(i,\mathcal{C},\mathcal{T})$ of \eqref{eq-IV 2 exam} into integers $n\in[N]=[12]$ according to the one-to-one mapping $\varphi$ in Table~\ref{tab-mapping}, we have the results as illustrated in \eqref{eq-IVs 2 nodes}.
\end{example}

Next, we propose the transmission strategy for the requested IVs belonging to Type II and Type III, to detail the two-rounds Shuffle and reduce phases.
In round 1, we focus on the IVs requested by $s-1$ nodes belonging to Type II.
For the nodes $\mathcal{K}=\mathcal{G}_i[\mathcal{C}]$, where $i\in[s],\mathcal{C}\in{[\Lambda]\choose t+1}$, each of which can multicast the following message,
\begin{align}
X_{\mathcal{K}}=\left(\bigoplus_{q\in\mathcal{C}} v_{q,(i,\mathcal{C},\mathcal{T})} \ : \ \mathcal{T}=\mathcal{C}\setminus\{q\}\right), \label{eq-X1-sum}
\end{align}
to the nodes in  $\{\mathcal{A}_q\setminus \mathcal{K}:q\in\mathcal{C}\}$ over the broadcast channel. Thus the number of nodes served by one message is
\begin{align}
g_1=|\mathcal{A}_q\setminus \mathcal{K}|\cdot|\mathcal{C}|=(s-1)(t+1). \label{eq-g1}
\end{align}
{ To ensure that each node contributes equally to a multicast message}, each IV included in \eqref{eq-X1-sum} is divided into $|\mathcal{K}|=t+1$ disjoint equal-size packets. For each node $k=\mathcal{K}[j],j\in[t+1]$, it multicasts the following message,
\begin{align}
X_k=X_{\mathcal{K}[j]}=\left(\bigoplus_{q\in\mathcal{C}} v_{q,(i,\mathcal{C},\mathcal{T})}^{(j)} \ : \ \mathcal{T}=\mathcal{C}\setminus\{q\}\right), \label{eq-X1}
\end{align}
to the nodes belonged to $\{\mathcal{A}_q\setminus \mathcal{K}:q\in\mathcal{C}\}$.
Furthermore, there are totally $S_1$ messages transmitted in round 1, where
\begin{align}
S_1=(t+1)s{\Lambda\choose t+1}. \label{eq-S1}
\end{align}
We can check that in round 1, there are totally multicast
\begin{align}
	\frac{g_1\cdot S_1}{(t+1)(s-1)}=\frac{(s-1)(t+1)\cdot(t+1)s{\Lambda\choose t+1}}{(t+1)(s-1)}=s(t+1){\Lambda\choose t+1}=N_1
\end{align}
distinct IVs, which matches $F_1$ in \eqref{eq-F1}, thus the decodability of round 1 is guaranteed.
\begin{example}
We return to the example in Section \ref{sec-example} with $K=6$, $r=3$, and $s=2$.
From \eqref{eq-X1-sum}, we have the following multicast messages in round 1 (e.g., message $X_{\{1,2\}}$ means that it is transmitted by nodes $1$ and $2$).
\begin{align}
X_{\{1,2\}}=&\left(\bigoplus_{q\in\mathcal{C}} v_{q,(1,\mathcal{C},\mathcal{T})}:  \mathcal{C}=\{1,2\},\mathcal{T}=\mathcal{C}\setminus\{q\}\right) \nonumber \\
=&v_{1,(1,\{1,2\},\{2\})}\bigoplus v_{2,(1,\{1,2\},\{1\})}\overset{\text{Table \ref{tab-mapping}}}{=}v_{1,2}\bigoplus v_{2,1}, \nonumber \\
X_{\{1,3\}}=&\left(\bigoplus_{q\in\mathcal{C}} v_{q,(1,\mathcal{C},\mathcal{T})}  :  \mathcal{C}=\{1,3\},\mathcal{T}=\mathcal{C}\setminus\{q\}\right) \nonumber \\
=&v_{1,(1,\{1,3\},\{3\})}\bigoplus v_{3,(1,\{1,3\},\{1\})}\overset{\text{Table \ref{tab-mapping}}}{=}v_{1,4}\bigoplus v_{3,3}, \nonumber \\
X_{\{2,3\}}=&\left(\bigoplus_{q\in\mathcal{C}} v_{q,(1,\mathcal{C},\mathcal{T})}  :  \mathcal{C}=\{2,3\},\mathcal{T}=\mathcal{C}\setminus\{q\}\right) \nonumber \\
=&v_{2,(1,\{2,3\},\{3\})}\bigoplus v_{3,(1,\{2,3\},\{2\})}\overset{\text{Table \ref{tab-mapping}}}{=}v_{2,6}\bigoplus v_{3,5}, \nonumber \\
X_{\{4,5\}}=&\left(\bigoplus_{q\in\mathcal{C}} v_{q,(2,\mathcal{C},\mathcal{T})}  :  \mathcal{C}=\{1,2\},\mathcal{T}=\mathcal{C}\setminus\{q\}\right) \nonumber \\
=&v_{1,(2,\{1,2\},\{2\})}\bigoplus v_{2,(2,\{1,2\},\{1\})}\overset{\text{Table \ref{tab-mapping}}}{=}v_{1,8}\bigoplus v_{2,7}, \nonumber \\
X_{\{4,6\}}=&\left(\bigoplus_{q\in\mathcal{C}} v_{q,(2,\mathcal{C},\mathcal{T})}  :  \mathcal{C}=\{1,3\},\mathcal{T}=\mathcal{C}\setminus\{q\}\right) \nonumber \\
=&v_{1,(2,\{1,3\},\{3\})}\bigoplus v_{3,(2,\{1,3\},\{1\})}\overset{\text{Table \ref{tab-mapping}}}{=}v_{1,10}\bigoplus v_{3,9}, \nonumber \\
X_{\{5,6\}}=&\left(\bigoplus_{q\in\mathcal{C}} v_{q,(2,\mathcal{C},\mathcal{T})}  :  \mathcal{C}=\{2,3\},\mathcal{T}=\mathcal{C}\setminus\{q\}\right) \nonumber \\
=&v_{2,(2,\{2,3\},\{3\})}\bigoplus v_{3,(2,\{2,3\},\{2\})}\overset{\text{Table \ref{tab-mapping}}}{=}v_{2,12}\bigoplus v_{3,11}.  \label{eq-X1 sum exam}
\end{align}
Notice that,  the involved IVs exactly coincide with the Type II IVs in \eqref{eq-IV 1 exam} and \eqref{eq-IVs 1 node}.

To be average, each IV included in \eqref{eq-X1 sum exam} is divided into $|\mathcal{K}|=2$ disjoint equal-size packets, and one of each is transmitted by one node belonging to  $\mathcal{K}$, as shown in \eqref{eq-X1}.
We can check that, the resulting multicast messages for each node $k\in[6]$ coincide with the $6$th column of Table~\ref{tab-exam all}, i.e., ``round I".
\end{example}

In round 2, we focus on the IVs requested by $s$ nodes  belonging to Type III.  For each node $k=\mathcal{G}_i[j], i\in[s], j\in[\Lambda]$, it multicast the following message,
\begin{align}
X_k=X_{\mathcal{G}_i[j]}=\left(\bigoplus_{q\in\mathcal{L}}v_{q,(i,\mathcal{C},\mathcal{T})} \ : \ \mathcal{L}\subseteq[\Lambda]\setminus\{j\},|\mathcal{L}|=t+1,\mathcal{C}=\{j\}\cup\mathcal{L}\setminus\{q\},\mathcal{T}=\mathcal{L}\setminus\{q\}\right), \label{eq-X2}
\end{align}
to the nodes belonging to $\{\mathcal{A}_q:q\in\mathcal{L}\}$ over the broadcast channel. Thus the number of nodes served by one message is
\begin{align}
g_2=|\mathcal{A}_q|\cdot|\mathcal{L}|=s(t+1)=r+s-1. \label{eq-g2}
\end{align}
Furthermore, there are totally $S_2$ messages transmitted in round 2, where
\begin{align}
	S_2=K{\Lambda-1\choose t+1}. \label{eq-S2}
\end{align}
We can check that in round 2, there are totally multicast
\begin{align}
\frac{g_2\cdot S_2}{s}&=\frac{s(t+1)\cdot K{\Lambda-1\choose t+1}}{s}=(t+1)\cdot s\Lambda\frac{\Lambda-(t+1)}{\Lambda}{\Lambda\choose t+1} \nonumber \\
&=s(t+1){\Lambda\choose t+1}(\Lambda-(t+1))=N_1(\Lambda-(t+1))
\end{align}
distinct IVs, which matches $F_2$ in \eqref{eq-F2}, thus the decodability of round 2 is guaranteed.
\begin{example}
We return to the example in Section \ref{sec-example} with $K=6$, $r=3$, and $s=2$.
From \eqref{eq-X2}, we have the following multicast messages in round 2.
\begin{align}
X_1=X_{\mathcal{G}_1[1]}&=\left(\bigoplus_{q\in\mathcal{L}}v_{q,(1,\mathcal{C},\mathcal{T})}:  \mathcal{L}=\{2,3\},\mathcal{C}=\{1,2,3\}\setminus\{q\},\mathcal{T}=\mathcal{L}\setminus\{q\}\right) \nonumber \\
&=v_{2,(1,\{1,3\},\{3\})}\bigoplus v_{3,(1,\{1,2\},\{2\})}\overset{\text{Table \ref{tab-mapping}}}{=}v_{2,4}\bigoplus v_{3,2}, \nonumber \\
X_2=X_{\mathcal{G}_1[2]}&=\left(\bigoplus_{q\in\mathcal{L}}v_{q,(1,\mathcal{C},\mathcal{T})}:  \mathcal{L}=\{1,3\},\mathcal{C}=\{1,2,3\}\setminus\{q\},\mathcal{T}=\mathcal{L}\setminus\{q\}\right) \nonumber \\
&=v_{1,(1,\{2,3\},\{3\})}\bigoplus v_{3,(1,\{1,2\},\{1\})}\overset{\text{Table \ref{tab-mapping}}}{=}v_{1,6}\bigoplus v_{3,1}, \nonumber \\
X_3=X_{\mathcal{G}_1[3]}&=\left(\bigoplus_{q\in\mathcal{L}}v_{q,(1,\mathcal{C},\mathcal{T})}:  \mathcal{L}=\{1,2\},\mathcal{C}=\{1,2,3\}\setminus\{q\},\mathcal{T}=\mathcal{L}\setminus\{q\}\right) \nonumber \\
&=v_{1,(1,\{2,3\},\{2\})}\bigoplus v_{2,(1,\{1,3\},\{1\})}\overset{\text{Table \ref{tab-mapping}}}{=}v_{1,5}\bigoplus v_{2,3}, \nonumber \\
X_4=X_{\mathcal{G}_2[1]}&=\left(\bigoplus_{q\in\mathcal{L}}v_{q,(2,\mathcal{C},\mathcal{T})}:  \mathcal{L}=\{2,3\},\mathcal{C}=\{1,2,3\}\setminus\{q\},\mathcal{T}=\mathcal{L}\setminus\{q\}\right) \nonumber \\
&=v_{2,(2,\{1,3\},\{3\})}\bigoplus v_{3,(2,\{1,2\},\{2\})}\overset{\text{Table \ref{tab-mapping}}}{=}v_{2,10}\bigoplus v_{3,8}, \nonumber \\
X_5=X_{\mathcal{G}_2[2]}&=\left(\bigoplus_{q\in\mathcal{L}}v_{q,(2,\mathcal{C},\mathcal{T})}:  \mathcal{L}=\{1,3\},\mathcal{C}=\{1,2,3\}\setminus\{q\},\mathcal{T}=\mathcal{L}\setminus\{q\}\right) \nonumber \\
&=v_{1,(2,\{2,3\},\{3\})}\bigoplus v_{3,(2,\{1,2\},\{1\})}\overset{\text{Table \ref{tab-mapping}}}{=}v_{1,12}\bigoplus v_{3,7}, \nonumber \\
X_6=X_{\mathcal{G}_2[3]}&=\left(\bigoplus_{q\in\mathcal{L}}v_{q,(2,\mathcal{C},\mathcal{T})}:  \mathcal{L}=\{1,2\},\mathcal{C}=\{1,2,3\}\setminus\{q\},\mathcal{T}=\mathcal{L}\setminus\{q\}\right) \nonumber \\
&=v_{1,(2,\{2,3\},\{2\})}\bigoplus v_{2,(2,\{1,3\},\{1\})}\overset{\text{Table \ref{tab-mapping}}}{=}v_{1,11}\bigoplus v_{2,9}. \label{eq-X2 exam}
\end{align}
Notice that,  the involved IVs exactly coincide with the Type III IVs in \eqref{eq-IV 2 exam} and \eqref{eq-IVs 2 nodes}.
We can check that, the resulting multicast messages for each node $k\in[6]$ coincide with the $7$th column of Table~\ref{tab-exam all}, i.e., ``round II".

\end{example}

\subsection{Communication Load}
From \eqref{eq-N}, \eqref{eq-Q}, \eqref{eq-S1}, and \eqref{eq-S2}, the communication load guaranteed by the scheme described above is given by
\begin{align}
L(r,s)&=\frac{S_1\cdot\frac{N}{N_1}\cdot\frac{Q}{Q_1}\cdot\frac{T}{t+1}+S_2\cdot\frac{N}{N_1}\cdot\frac{Q}{Q_1}\cdot T}{NQT}=\frac{S_1/(t+1)+S_2}{N_1Q_1} \\
&=\frac{s{\Lambda\choose t+1}+K{\Lambda-1\choose t+1}}{\Lambda s(t+1){\Lambda\choose t+1}} =\frac{s{\Lambda\choose t+1}+\frac{K(\Lambda-(t+1))}{\Lambda}{\Lambda\choose  t+1}}{\Lambda s(t+1){\Lambda\choose t+1}} \\
&=\frac{\Lambda-t}{\Lambda(t+1)}=\frac{K/s-(r-1)/s}{K/s((r-1)/s+1)} \\
&=\frac{s(K-r+1)}{K(r+s-1)}, \label{eq-L}
\end{align}
which coincides with Theorem~\ref{th-proposed scheme}.
\begin{remark}
When $s=1$, the communication load guaranteed by the scheme described above is given by
\begin{align}
L(r)=\frac{S_2\cdot\frac{N}{N_1'}\cdot\frac{Q}{Q_1'}\cdot T}{NQT}=\frac{S_2}{N_1'Q_1'}=\frac{K{K-1\choose r}}{Kr{K\choose r}}=\frac{1}{r}\cdot(1-\frac{r}{K}),
\end{align}
which is the same as the result \cite[Theorem 1]{LMYA}.
\end{remark}

\section{Conclusion}
\label{sec-conclusion}
In this paper, a novel cascaded CDC scheme is proposed by deliberately designing the data placement and output functions assignment strategy with a grouping method.
Based on it, we provided the two-rounds Shuffle strategy with the multicast gain $(r+s-1)(1-1/s)$ and $r+s-1$.
Under the proposed data placement and output functions assignment, the communication load $L$ of the proposed scheme was proved to be order optimal within a factor of $2$, and approximately optimal when $K$ is sufficiently large for a given $r$.
Compared with the well-know scheme in \cite{LMYA}, our scheme breaks the fundamental limits through new data placement and output function assignment strategies. We demonstrate that these limits can be surpassed even when $r\neq s$.
Compared with the low-complexity schemes supporting for flexible $K$ \cite{JQ,WCJ}, the proposed scheme has the potential to improve both the communication load and the required number of files.
Furthermore, when $s=1$, the result of the proposed scheme is the same as that of \cite{LMYA}, which means the optimality is guaranteed.
On-going works include the construction for a more general case of $K/s\notin\mathbb{N}$.

\appendices
\section{Proof of Theorem~\ref{th-converse}}
\label{app-converse}
From \cite[Proposition 2]{KNL}, we denote $P_1$, $P_2$ as the subsets of computing nodes, denote $b^{P_2}_{P_1}$ as the bits of IVs that are demanded exclusively by the nodes in $P_1$ and available in the nodes of $P_2$. Then, the converse bound  $L^{\star}$ is given by
\begin{align}
L^{\star}(r,s)\geq\frac{1}{QNT}\sum_{P_1\subseteq[K]}\sum_{P_2\subseteq[K]\setminus P_1}\frac{|P_1|}{|P_1|+|P_2|-1}b^{P_2}_{P_1}.\label{eq-L* general}
\end{align}

For the proposed scheme, we have $|P_2|=r$ since each IV is generated by $r$ times.
Assume that $|P_1|=l$, the possible values for $l$ belong to $\{s-1, \min\{K-r,s\}\}$, which  corresponded to the Type II IVs and Type III IVs, respectively.
From \eqref{eq-F1} and \eqref{eq-F2}, we have:
\begin{itemize}
	\item If $l=s-1$, then $\frac{1}{QNT}b^{P_2}_{P_1}=\frac{N_1}{Q_1N_1}$;
	\item If $l=\min\{K-r,s\}$, then  $\frac{1}{QNT}b^{P_2}_{P_1}=\frac{N_1(\Lambda-(t+1))}{Q_1N_1}$.
\end{itemize}
Thus, under the proposed strategies of data placement and output functions assignment, \eqref{eq-L* general} is rewritten as,
\begin{align}
L^{\star}(r,s)&\geq\frac{s-1}{s-1+r-1}\cdot\frac{N_1}{Q_1N_1}+\frac{s}{s+r-1}\cdot\frac{N_1(\Lambda-(t+1))}{Q_1N_1} \nonumber \\
&=\frac{s(s-1)}{K(r+s-2)}+\frac{s(K-(r+s-1))}{K(r+s-1)} \label{eq-L*}
\end{align}
By adjusting the first term of \eqref{eq-L*}, we have
\begin{align}
L^{\star}(r,s)&>\frac{s(s-1)}{K(r+s-1)}+\frac{s(K-(r+s-1))}{K(r+s-1)} \nonumber\\
&=\frac{s(K-r)}{K(r+s-1)}. \label{eq-L* proof}
\end{align}

From \eqref{eq-Load} in Theorem~\ref{th-proposed scheme} and \eqref{eq-L* proof}, we have,
\begin{align}
\frac{L}{L^{\star}}<\frac{s(K-r+1)}{K(r+s-1)}\cdot\frac{K(r+s-1)}{s(K-r)}=1+\frac{1}{K-r}\leq2.\label{eq-order optimal}
\end{align}
Thus, the proposed scheme is order optimal within a factor of $2$.
Furthermore, for a given $r$, when $K\to\infty$, we have $\frac{L}{L^{\star}}\approx1$, i.e., the proposed scheme is asymptotically optimal. Hence, we prove Theorem~\ref{th-converse}.

\bibliographystyle{IEEEtran}
\bibliography{ref}

\end{document}